\documentclass[twocolumn,prd,superscriptaddress,aps,nofootinbib,floats,floatfix,amsmath,amssymb,secnumarabic]{revtex4}
\usepackage[dvips,final]{graphicx}
\usepackage{amsmath}
\usepackage{amsfonts}
\usepackage{amssymb}
\usepackage{latexsym}
\usepackage{graphicx}
\usepackage{verbatim}
\usepackage[english]{babel}
\newcommand{\be}{\begin{equation}}
\newcommand{\ee}{\end{equation}}
\newcommand{\ba}{\begin{array}}
\newcommand{\ea}{\end{array}}
\newcommand{\bqa}{\begin{eqnarray}}
\newcommand{\eqa}{\end{eqnarray}}
\renewcommand{\d}{\mathrm{d}}

\newcommand{\vl}{\emph{Via Lactea II}}
\newcommand{\sv}{\langle \sigma v \rangle}
\newcommand{\bsh}{B_{\rm\scriptscriptstyle SH}}
\newcommand{\bmh}{B_{\rm\scriptscriptstyle MH}}
\newcommand{\mchi}{M_{\rm\scriptscriptstyle DM}}
\newcommand{\sfrac}[2]{\textstyle\frac{#1}{#2}}
\newcommand{\Egp}{E_{\gamma'}}
\newcommand{\Eg}{E_{\gamma}}
\newcommand{\epm}{{e^{\pm}}}
\newcommand{\lsim}{\mbox{\raisebox{-.6ex}{~$\stackrel{<}{\sim}$~}}}
\newcommand{\gsim}{\mbox{\raisebox{-.6ex}{~$\stackrel{>}{\sim}$~}}}

\begin{document}

\title{Overcoming Gamma Ray Constraints with Annihilating Dark Matter\\ in Milky Way Subhalos}

\author{Aaron C. Vincent\footnote{vincenta@hep.physics.mcgill.ca}}
\affiliation{Department of Physics, McGill University,
3600 Rue University, Montr\'eal, Qu\'ebec, Canada H3A 2T8}

\author{Wei Xue\footnote{xuewei@hep.physics.mcgill.ca}}
\affiliation{Department of Physics, McGill University,
3600 Rue University, Montr\'eal, Qu\'ebec, Canada H3A 2T8}

\author{James M. Cline\footnote{jcline@hep.physics.mcgill.ca}}
\affiliation{Department of Physics, McGill University,
3600 Rue University, Montr\'eal, Qu\'ebec, Canada H3A 2T8}
\affiliation{Perimeter Institute for Theoretical Physics,
31 Caroline St N, Waterloo, Ontario N2L2Y5, Canada}

\begin{abstract} 

We reconsider  Sommerfeld-enhanced annihilation of dark matter (DM) into leptons to
explain PAMELA and Fermi electron and positron observations, in light of possible new
effects from substructure.   There is strong tension between getting a large enough
lepton  signal while respecting constraints on the fluxes of associated gamma rays; we show how DM
annihilations within subhalos can get around these constraints.  Specifically, if most
of the observed lepton excess comes from annihilations in a nearby (within 2 kpc)
subhalo along a line of sight toward the galactic center, it is possible to match both
the lepton and gamma ray observations. We demonstrate that this can be achieved in a
simple class of particle physics models in which the DM annihilates via a hidden
leptophilic U(1) vector boson, with explicitly computed Sommerfeld enhancement factors.
Gamma ray constraints on the main halo annihilations (and CMB constraints from the era
of decoupling) require the annihilating component of the DM to be subdominant, of order
$10^{-2}-10^{-3}$ of the total DM density.
\end{abstract}
\maketitle
\newpage

\section{Introduction} Nongravitational signals of Dark Matter (DM) have been sought
after for some time now by the astrophysical and particle physics communities. At the
same time results from the Payload for Antimatter Matter Exploration and Light-nuclei
Astrophysics (PAMELA) experiment and from the Fermi space telescope suggest a local
excess positron fraction $e^+/(e^+ + e^-)$ at energies above 10 GeV as well as an
excess of $e^+ + e^-$ peaking around 500 GeV. Standard cosmic ray propagation models do
not account for these excesses. An attractive explanation is that a DM WIMP (weakly
interacting massive particle) is present
in our galaxy at large enough concentrations to self-annihilate into standard model
leptons. A TeV-scale WIMP annihilating to electron-positron pairs could produce
such signals. In order to be consistent with the observed relic abundance of DM, the
annihilation cross-section $\sv_0 \sim 3 \times 10^{-26}$ cm$^3$ s$^{-1}$ would have to
be enhanced by a factor of order 100, for example by a velocity-dependent
Sommerfeld enhancement.

Many authors
\cite{Cirelli:2008pk,ArkaniHamed:2008qn,Pospelov:2008jd,Cholis:2008qq,Cholis:2008wq,Meade,Chen:2009ab,
Watson, Kuhlen:2008aw,Feng:2010zp}  have explored this
possibility, and have constrained the allowable mass versus boost factor parameter
space.   However these papers assume that the dominant source of indirect signals is
from annihilations in the main DM halo. In a previous work \cite{Cline:2010ag} we
considered the possibility adding the effects of dark matter substructure to the
theoretical model and we found examples where annihilations in subhalos could provide a
significant fraction of the observed lepton excesses.  We showed that one could find a
better overall fit to the electron-positron data from the Fermi and PAMELA experiments,
and we suggested that gamma ray constraints which are now putting considerable pressure 
on these
models could be alleviated.  Our purpose in the present work was to ascertain whether
this is indeed the case.

The constraints mentioned come from recent gamma ray observations of
the galaxy and from Cosmic Microwave Background (CMB) measurements.
As high energy electron-positron pairs are produced and diffuse
throughout the galaxy, they will emit final-state radiation as well
as scatter on the ambient photon field, giving rise to $\sim$ 1-100
GeV gamma rays that should be detectable. Given the large expected
concentrations of both DM and radiation near the galactic center
(GC), gamma rays from inverse Compton scattering (ICS) near the GC
are particularly constraining. The Fermi Large Area Telescope (LAT)
is specifically designed to detect gamma rays in this range, and its
latest results have been used to rule out large regions of parameter
space for annihilating WIMP models
\cite{Meade,Cirelli:2009dv,Strumia:2010zz,Cirelli:2010nh}. 

However in this work we will show that if a sizeable proportion of the leptons from DM
annihilation originate from nearby subhalos, the constraints from GC gamma rays can be
relieved. Final-state (bremsstrahlung) radiation from subhalos has been examined by
other authors \cite{Bovy,Kuhlen:2009jv,Kistler:2009xf,Ando:2009fp,
kuhlen,Kuhlen:2008aw,Hutsi:2010ai}, and ref.\ \cite{Kuhlen:2009is} has studied the $e^++e^-$
spectrum from a nearby subhalo.  In this follow-up work we extend our previous findings
to a prediction of the gamma ray spectrum including a full calculation of ICS radiation
in the galaxy, which we compare to the full-sky data from the Fermi LAT.  We include
the expected contribution to the gamma ray background coming from background electrons
and positrons. Using a fully-numerical approach, we find that there is less room for new contributions from the
annihilation products of the DM, making the constraints on the DM models more severe.
This is a serious issue even for less cuspy and cored DM profiles, that have been shown
to satisfy the constraints in previous semi-analytic treatments which ignored the background gamma
ray fluxes.

In our previous paper we focused on the contributions of distant subhalos to the flux
of leptons at Earth.  Even though these new contributions can improve the fit to the 
lepton data alone, here we show that they do not soften the gamma ray constraints
sufficiently to be viable.  Instead, we focus on the possibility  that an accidentally
nearby subhalo could provide the bulk of the leptonic flux.  The associated gamma rays
would be sufficiently hidden by strong backgrounds if this subhalo happened to lie
between us and the galactic center.  The effects of nearby subhalos have been
previously considered by ref.\ \cite{Brun:2009aj}, but only allowing for purely
astrophysical boost factors, due to the density of the subhalos.  Here
we find that velocity-dependent Sommerfeld enhancement is crucial for obtaining a 
positive outcome.  It is precisely because of the larger boost factor available within
subhalos (which have orders of magnitude smaller velocity dispersion) relative to the
main halo that we are able to soften the gamma ray constraint due to the main halo near
the GC, yet have a large enough lepton signal from a nearby subhalo. In addition, we
must assume that the leptophilic component of the DM responsible for these processes is
subdominant to the main inert (for our purposes) component, in order to sufficiently
reduce the effective boost factor for annihilations in the main halo
\cite{Cirelli:2010nh}.  This gives rise to the interesting possibility that different
kinds of DM are responsible for the cosmic ray anomalies than those which might
manifest themselves in direct detection experiments.

Using a modified version of the cosmic ray propagation code GALPROP and the data from
the recent \vl\ simulation of dark matter evolution and collapse in a Milky Way-sized
galaxy, we modelled the two-dimensional axisymmetric distribution of electrons and
positrons in the galaxy. These results were combined with simulated interstellar
radiation field (ISRF) data in order to compute a realistic skymap of the gamma ray
spectrum expected from DM annihilation in the Galaxy, which was in turn compared with a
year's worth of diffuse gamma ray observation from the Fermi LAT.

We start with a summary the cosmic ray model and results of our previous work in
Section \ref{sec:cr}, before discussing the relevant ICS and gamma ray physics in
Section \ref{sec:gr}.  In Section \ref{sec:results} we describe our methodology, and
present  model-independent fits to the data in several scenarios for the distribution
of subhalos and the halo profiles.    In particular, we show that an accidentally
nearby subhalo can provide a promising loophole to the gamma ray constraints on cuspy
profiles.  We also predict the gamma ray flux from the subhalo, which could provide a
test of the model if future measurements and understanding of backgrounds are improved.
 In section \ref{sec:particle} we then demonstrate that the boost factors
required for this scenario can be explicitly realized in a simple class of hidden
sector particle physics models.  We conclude  with a discussion of the overall
viability of this picture in section \ref{sec:conclusion}.

\section{Cosmic Ray Propagation}
\label{sec:cr}

Inside the galactic diffusion zone, particles and nuclei propagate according to the
diffusion-loss equation~\cite{Strong:1998pw}, which applies to electrons and positrons as
follows:\footnote{The full transport equation also includes the effects of convection and
diffusive reacceleration, which are mainly important for the propagation of heavier species.
Here we leave these terms out for clarity, although they were included in our full
calculations with GALPROP. These are important for determining the abundance of secondary
electrons and positrons, which come from spallation and decay of various species.}
\begin{eqnarray}
\frac{\d}{\d t}\psi_\epm(\textbf{x},\textbf{p},t) &=& Q_\epm(\textbf{x},E)+ \nabla 
\cdot\left(D(E) \nabla \psi_\epm(\textbf{x},\textbf{p},t)\right)   \nonumber \\
 &+& \frac{\partial }{\partial E} \left[b(\textbf{x},E)\psi_\pm(\textbf{x},\textbf{p},t) 
\right] \ .
\label{diffeq}
\end{eqnarray}
$\psi_\epm(\textbf{x},\textbf{p},t)$ denotes the particle number density per unit
momentum $|\textbf{p}|$, $Q$ represents the source function, $D(E)$ is the spatial
diffusion coefficient and $b(\textbf{x},E)$ is the energy loss coefficient. We seek the
steady-state solution of equation (\ref{diffeq}):
$\d\psi_{\epm}(\textbf{x},\textbf{p},t)/\d t=0$. 

Since (\ref{diffeq}) is linear, the leptons from DM annihilation travel independently
in the astrophysical background. The source  $Q_\epm$ comes from DM annihilation which
depends on the particle physics and the local density of the dark matter:
\begin{equation}
Q_{\epm}= \frac{1}{2} \left(\frac{\rho(\textbf{x})}{M} \right)^2 \sv 
\frac{\d N_{\epm}}{\d E} =\frac{n_{DM}^2}{2} BF \sv_0  \frac{\d N_{\epm}}{\d E}.
\label{source} \ , 
\end{equation}
where the prefactor $1/2$ is a symmetry factor for self-annihilation,
$n_{DM}(\textbf{x},E)$ is the DM energy density, $\sv_0= 3 \times 10^{-26 }$ cm$^{3}$
s$^{-1}$ is the benchmark value for standard cosmology to explain the relic density of
DM, and $\d N_{\epm}/\d E$ is the energy spectrum of the annihilation products.
Neglecting the effect of soft photons, the spectrum can be approximated by the simple
form $\d N_{\epm}/\d E = 2 M_{DM}^{-1}\Theta(M_{DM}-E)$, where $\Theta(x)$ is the usual
Heaviside step function, and the factor $2$ arises because that the final state has two
electrons or two positrons. The latter has the correct qualitative shape, and is easier
to implement in GALPROP than would be a more exact spectrum. $BF$ denotes the boost
factor due to Sommerfeld enhancement,  originating from a nonperturbative $\sim 1/v$
correction
due to the slow ($v/c<\alpha$) motion of the DM particles. 

To simplify our analysis, we take the boost factor $BF$ to be constant throughout the
main halo, and tune it to provide the best possible fit to available electron and
positron data. Since the Sommerfeld effect depends strongly on velocity,  typical
subhalos, which have a much smaller velocity dispersion, have a much higher $BF$, and
we treat it as an additional free parameter.  Although each subhalo has different
values of $BF$, we represent the subhalo $BF$ by a single average value in this first
part of our analysis, where the $BF$s are treated as being uncorrelated and best fit
values are sought.  This is not a limitation in the case we will eventually focus upon,
namely domination of the excess lepton signal by a single nearby subhalo.  A further
complication is that in fact $BF$ has a radial dependence within each halo, because the
velocity dispersion is a function of $r$, which has been fitted by   many-body
simulations such as \vl\ \cite{kuhlen}.  We will take this into account in section
\ref{bavg} by averaging $BF$ over the phase space of DM in the halos, in order to make
contact with the results obtained in this model-independent part of our analysis.

The spatial diffusion coefficient can be parametrized as follows~\cite{Simet:2009ne}:
\begin{equation}
D(E)=D_0 \left(\frac{E}{4 \ \text{GeV}}\right)^\delta
\label{diffusioncoefficient}
\end{equation}
 Two widely-used approaches exist for solving the diffusion equation in the Galaxy:
semianalytic and fully numerical. We chose the latter for Galaxy-scale propagation, in
part because a numerical approach allows for better control over the spatial dependence
of the astrophysical input, such as energy loss due to inverse Compton scattering.
GALPROP 50.1p~\cite{GALPROP:web} is a publicly available software package that solves
Eq.\ (\ref{diffeq}) with an implicit-in-time 2D or 3D Crank-Nicholson scheme. In 2D
mode, it provides a ($r,z$) map in cylindrical coordinates of the number density of
each species within the Galactic diffusion zone. To constrain the diffusion parameters,
the ratio of measured  secondary-to-primary species such as B/C or sub-Fe/Fe can be
simulated and fit to observations. This was done to a very high degree of accuracy in
Ref.\ \cite{Simet:2009ne}. We used results from their best fits: $D_0=6.04 \times
10^{28}\, \text{cm}^2 \text{s}^{-1} (0.19\, \text{kpc}^2/\text{Myr})$, and $\delta=0.41$. 

\begin{figure}
\hspace{-0.4cm}
\includegraphics[width=0.7\textwidth,angle=90]{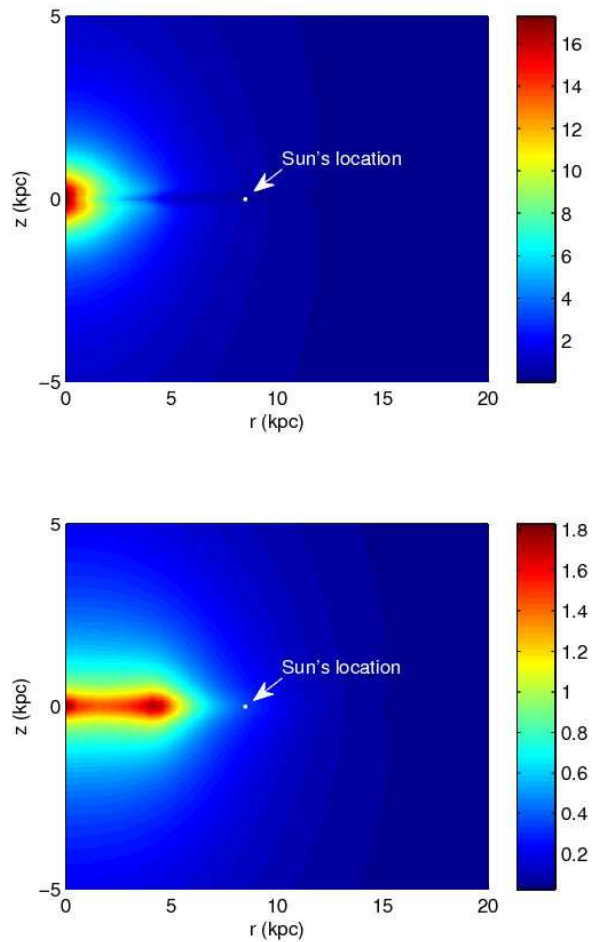}
\caption{Simulated energy density distribution of the interstellar radiation field (ISRF)
within the Milky Way by \cite{Porter:2005qx}, integrated over energies. Top: starlight
component. Bottom: IR component, from dust. The CMB component is of course uniform
throughout the galaxy. Color scale is $log(density)$ in arbitrary units.}
\label{ISRFfig}
\end{figure}

The full energy loss rate is due to synchrotron radiation and inverse Compton
scattering: 
\begin{equation}
b(x,E)=-\frac{\d E_e}{\d t} = \frac{32 \pi \alpha_{em}}{3 m_e^4} E_e^2 \left[ u_B + 
\sum_{i=1}^3 u_{\gamma i} \cdot R_i(E_e)     \right] \ .
\end{equation}
$\alpha_{em}$ is the fine structure constant and  $u_B= B^2/2$ is the energy density of the
galactic magnetic field, for which we used the standard parametrization:
\begin{equation}
B(r,z) \simeq 11 \mu G \cdot \text{exp}\left( -{r\over 10\, \text{kpc}} 
-{|z|\over 2\, \text{kpc} }\right).
\end{equation}
$u_{\gamma i}$ are the energy densities of the three main components of the
interstellar radiation field (ISRF): CMB radiation, thermal radiation from dust and
starlight, which lie mainly in the microwave, infrared and optical regions of the
electromagnetic spectrum, respectively. GALPROP uses position-dependent maps of ISRF
compiled by \cite{Porter:2005qx}, rather than using a constant energy-loss coefficient
computed from a local average. The latter approach (explained in section 3 of
\cite{Delahaye:2008ua}) is commonly used in the semi-analytic model.  While it is
indeed quite accurate when dealing with electrons from a smooth Galaxy-wide
distribution of dark matter, it is an approximation that is less precise when
considering the propagation into the Galaxy of electrons from DM subhalos outside of
the diffusion zone. We will nonetheless make use of the semianalytic method in Section
\ref{sec:closeSub}, when only local propagation will be relevant. The position
dependence of the ISRF in the Galaxy is presented in Figure \ref{ISRFfig}. Further
details will be discussed in section \ref{sec:ICS}. 

\subsection{Via Lactea II and GALPROP}
\label{sec:simulations}

We assumed that the DM was composed of a single Dirac fermion $\chi$ of mass
$M_{DM}$ annihilating through the channel $\chi \chi \rightarrow B B$, followed by the
decay $B \rightarrow e^+e^-$, where $B$ is some dark sector gauge boson which could
also be responsible for the Sommerfeld enhancement. We considered two astrophysical models
for the DM distribution: a main halo-only (MH) scenario, in which only a large,
spherical halo contributed annihilation products; and a subhalo (MH+SH) scenario, where
the overdensities formed by DM substructure were responsible for extra annihilation of
DM into electrons and positrons. In both cases, we used a spherically symmetric Einasto
profile for the DM density distribution:
\begin{equation}
\rho_{\rm Ein}(r)=\rho_s\, \text{exp} \left\{ -\frac{2}{\alpha} \left[ \left(\frac{r}{r_s}
\right)^{\alpha}-1 \right] \right\} \ .
\label{EinastoEq}
\end{equation}
$r$ is the radial coordinate from the center of the halo, $\rho_s$ is the density at $r
= r_s$, the distance at which the slope $\rho'/\rho = -2$. These parameters are simply
related to the radius and rotational velocity of a given subhalo as explained in Ref.\
\cite{kuhlen}. The shape parameter $\alpha$ can be read off from curve-fitting the
distributions from $N$-body simulations such as \cite{vl2, aquarius}. It is generally
taken to be around $\alpha \simeq 0.17$. We took $r_s = 25$ kpc for the main galactic
halo, with a local dark matter density $\rho_\odot = 0.37$ GeV cm$^{-3}$ in
agreement with \vl\ and with other recent estimates, {\it e.g.,} \cite{Catena:2009mf}. 
It should be noted that many authors use the convention $\rho_\odot =
0.3$ GeV cm$^{-3}$. This leads to a factor of 
$(0.3/0.37)^2 = 0.66$ difference in the constraints on the
annihilation cross sections, but it is 
of no consequence when it comes to excluding models, 
since constraints come from the ratio of gamma rays lepton fluxes, 
which both scale linearly with $\rho_\odot^2\sv$. 

It has been argued that direct observations of rotation velocities in the Milky Way
are consistent with cored DM profiles (see for example ref.\ \cite{Salucci:2010pz}).
Two such examples are the isothermal and Burkert \cite{Salucci:2000ps} ansatzes.
The Burkert profile has been fitted to the rotation curves of galaxies other than our
own, but we are not aware of references which attempt to fit the Milky Way.  To allow
for the alternative possibility of a cored main halo, we will
therefore restrict our attention to the isothermal profile
\be
	\rho_{\rm iso}(r) = {\rho_s\over 1 + (r/r_s)^2}
\ee
adopting the values $r_s = 3.2$ kpc and $\rho_s = 3.0$ GeV/cm$^3$ similar to those
used by ref.\ \cite{Cirelli:2009dv}.  These values are motivated by the
constraint on the observed solar density $\rho_\odot$ (which we take to be somewhat 
higher than in \cite{Cirelli:2009dv}) and on the mass of the Galaxy
with 50 kpc as determined from circular velocity measurements.  However for the subhalos
we will in all cases assume the Einasto form that is suggested by \vl.

\vl\ \cite{vl2} was a billion-particle simulation that tracked the evolution and collapse of
$10^9$ particles over the history of a Milky Way-sized structure. Data about the main
galactic halo and the 20,047 largest subhalos that the  particles (each taken to have
mass 4,100 $M_\odot$)
merged into over the course of the simulation are available to the public. While the
visible galaxy is only 40 kpc across, these subhalos extend as far out as 4000 kpc from
the GC. We used the \vl\ subhalo data as a model for substructure  sourcing 
electrons and positrons (from DM annihilation) at the
boundary of the GALPROP diffusion zone, with an overall tunable boost factor for the
subhalo annihilation rate. In addition to a larger Sommerfeld enhancement from smaller
velocity dispersions within each subhalo, we expect sub-substructure unresolvable from
numerical simulations to give rise to further enhancement of the annihilation
cross-section.  Recent estimates \cite{Kuhlen:2008aw} show that such sub-subhalos alone
could increase annihilation rates by as much as a factor of 10. 

\begin{figure}
\hspace{-0.4cm}
\includegraphics[width=0.6\textwidth,angle=90]{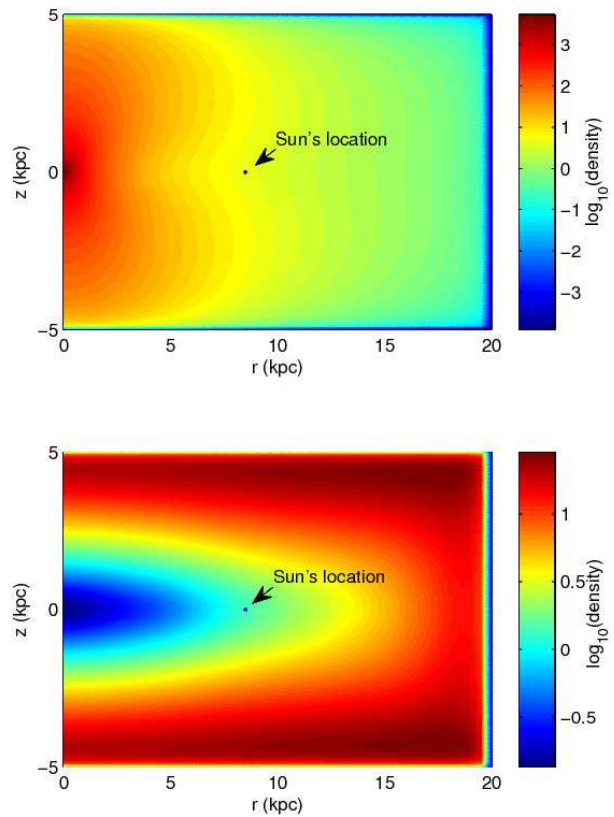}
\caption{Simulated steady-state distribution of electrons and positrons from DM annihilation
within the Milky Way diffusion zone. The galactic center is located at $z = 0$, $r = 0$; red
corresponds to high densities, blue to low densities. Top: leptons from the main halo only.
Bottom: leptons from the subhalos only, sourced from the diffusion zone boundary. Note that
the scales are different: the peak main halo density (at the GC) is about 200 times larger
than the peak subhalo density (near the edge of the diffusion zone)}
\label{electronFigures}
\end{figure}

Electrons from an extragalactic source have a very particular density profile. While
the annihilation products from the main halo follow a roughly symmetric distribution
about the GC, SH electrons sourced from the diffusion zone boundary tend to form a
diffuse ``shell'' near the edge of the diffusion zone, as illustrated in Fig.\ 
\ref{electronFigures}. Ambient radiation prevents high-energy particles from reaching
the GC, trapping them near the edge of the Galaxy. The large number of subhalos
combined with a large boost factor can allow some particles to make their way to
earth, albeit with a fraction of their initial energy.

We compared the best-fit combination of DM mass and boost factor for the MH scenario
with the best fits for the MH+SH scenario in \cite{Cline:2010ag}. The results  are
summarized in table \ref{resultTable}:  a much better fit could be obtained by
including subhalos and a dark matter particle with $M_{DM} = 2.2$ TeV, rather than the
standard MH-only $M_{DM} = 1$ TeV.   Of course, the fits are further  improved by
allowing the normalizations of the background electrons and positrons to be additional
free parameters, denoted as the ``freely varying background,'' as opposed to the
standard backgrounds resulting from GALPROP simulations which include the effects of heavier nuclear species.  Assuming this extra freedom has been
advocated or used by numerous authors \cite{Meade,Cirelli:2009dv,
Cholis:2008wq,Papucci:2009gd}.  In table \ref{resultTable} we also show the fit we
obtain in the present analysis for the main-halo-only case with an isothermal profile
and fixed background.  It is significantly worse than the corresponding one for an
Einasto profile.

\subsection{Annihilation channels}
\label{sec:channels}

While we have mostly focused on the 4e final state, there is no
reason for other, heavier particles not to be produced if the mass of
the intermediate gauge boson is large enough. Since the amount of
Sommerfeld enhancement ultimately depends on this mass, it is
important to include the decays to muons and pions.  The possible
final states are all the four-particle combinations of 
$2e$, $2\mu$ and $2\pi$. The muon and pion spectra are given by Ref.\
\cite{Cholis:2008vb}, whose authors were kind enough to provide us
with the appropriate GALPROP implementation. 

The branching ratios are given by $r_i = f_i/\sum f_i$, where
the $f_i$ are given by
\be
	f_i = \sqrt{\mu^2-4 m_i^2}\left\{\begin{array}{ll}
	4(\mu^2 + 2 m_i^2),& i=e,\mu\\
	(\mu^2 - 4 m_i^2),&  i=\pi\end{array}\right.
\label{branchratios}
\ee
In each $f_i$, the square root factor comes from the phase space,
while the rest is from the squared matrix element for the decay. 
Below threshold, $f_i$ is defined to be zero.  For a gauge boson with
a mass $\mu \gsim 1$ GeV, we find $r_e = r_\mu = 0.45$ and $r_\pi =
0.1$. In this case the electrons produced from the final decay of the
$\mu$'s and $\pi$'s peak at a lower energy, thus requiring a slightly
higher mass of $M_{DM} = 1.2$ TeV in order to fit the Fermi and
PAMELA data. This is much smaller than the well-known $M_{DM} \simeq$
2.2 TeV best fit in the pure-muon final state
\cite{Meade,Papucci:2009gd,Cholis:2008wq} because of the large
fraction of gauge bosons still decaying directly to high-energy
electrons. These results are also shown in Table \ref{resultTable}.


\begin{table}
 \begin{tabular}{|l|c|l|l|l|l|l|}
\hline
 \multicolumn{7}{|c|}{Freely-varying background (Einasto)} \\
 \hline
 & $\mchi$ (TeV) & $\chi^2_{\rm Fermi}$ &
$\chi^2_{\rm\scriptscriptstyle PAMELA}$ & $\chi^2_{\rm total}
$&$\bmh$& $\bsh$ \\ \hline
MH ($4e$) & 0.85 & 15.5 & 18.7 & 34.3 & 90.3 & \ \ $-$\\
MH+SH & 1.2 & 2.3 & 14.2 & 16.5 &92.8 & 3774 \\
\hline
\multicolumn{7}{|c|}{Fixed GALPROP background (Einasto)} \\
\hline
MH ($4e$)               & 1.0 & 8.2 & 144 & 152 & 110 & \ \ $-$\\
MH+SH           & 2.2 & 2.1 & 175 & 177 & 146 & 1946 \\
MH ($e,\mu,\pi$)        & 1.2 & 3.8 & 109 & 112 & 118 & \ \ $-$ \\
\hline
\multicolumn{7}{|c|}{Isothermal profile (fixed background)} \\
\hline
MH ($4e$)               & 1.0 & 9.1 & 186 & 195 & 113 & \ \ $-$ \\
MH ($e,\mu,\pi$)        & 1.2 & 3.0 & 151 & 154 & 119 & \ \ $-$ \\
\hline
\end{tabular}
\caption{First four rows: best fit results from \cite{Cline:2010ag}, assuming
Einasto profile. By varying the boost factors of 
the main halo and faraway subhalos separately, we found that the fit to the 
PAMELA and Fermi data from MH annihilations alone could be improved by
inclusion of SH annihilations as shown. 
Last two rows: new fit for isothermal profile ($r_s=3.2$ kpc, $\rho_s = 3.0$ GeV/cm$^3$),
main-halo-only scenario from this work, using the fixed GALPROP
background, and same parameters as in \cite{Cline:2010ag}. 
We assume the annilation to the 4e final state, except in the cases
``MH ($e,\mu,\pi$)'' which indicates the the process $\chi \chi
\rightarrow B B \rightarrow 4\ell$, where $\ell$ stands for $e^\pm$,
$\mu^\pm$ or $\pi^\pm$, with branching ratios  $r_e = r_\mu = 0.45$
and $r_\pi = 0.1$ as explained in Section \ref{sec:channels}.}
\label{resultTable} \end{table}

\section{Gamma Ray Computation from Inverse Compton Scattering and Bremsstrahlung}
\label{sec:gr}
\subsection{``Prompt'' gamma ray emission (bremsstrahlung)}

Prompt gamma ray emission appears in the final stage of DM annihilation, softening the
lepton spectrum. 
The flux can be divided into main halo 
and subhalo parts:
\begin{equation}
 \frac{\d\Phi}{\d \Eg \d \Omega} = \frac{\d\Phi_{\rm main}}{\d \Eg \d \Omega} + 
\frac{\d\Phi_{\rm sub}}{\d \Eg \d \Omega}  \ .
\end{equation}
The astrophysical and particle physics dependences of each flux can be factorized
as
\begin{equation}
 \frac{\d\Phi_{\rm main}}{\d \Eg \d \Omega} = \frac{1}{2} 
\frac{\sv}{4 \pi} r_{\odot} \frac{\rho_{\odot}^2}{m_{\chi}^2} 
\frac{\d N}{\d \Eg} \bar{J}_{main}
\label{phiMH}
\end{equation}
and
\begin{equation}
  \frac{\d\Phi_{\rm sub}}{\d \Eg \d \Omega} = \frac{1}{2} \sv\frac{\d N}{\d \Eg} 
\bar{J}_{sub}.
\label{phiSH}
\end{equation}
In each case, the $\bar{J}_i$ factor depends only upon astrophysical
inputs. The main halo $J$ factor is defined as a line of sight (l.o.s.) integral of flux at 
each pixel:
\begin{equation}
\bar{J}_{\rm main} = \frac{1}{\Delta \Omega}\int_{\Delta \Omega} \d \Omega \int_{\rm l.o.s.}
  \frac{\d s}{r_{\odot}}
  \left( \frac{ \rho_{\rm main}[r(s,\psi)] }{\rho_{\odot}} \right)^2.
\label{JMH}
\end{equation}
In the case of flux originating from many distant subhalos, we may treat each
one as a point source of radiation. In this case, the diffuse
flux per solid angle requires a sum over each contributing source with density $\rho_i$
and distance $d_i$ within the observed solid angular region $\Delta \Omega$:
\begin{equation}
\bar{J}_{\rm sub}  = \frac{1}{\Delta \Omega}\sum_{\Delta\Omega} \left( \frac{1}{4 \pi d_i^2}
  \int \d V \frac{\rho_i^2}{m_{\chi}^2} \right).
\end{equation}
This clearly depends not only on the density profiles, but also on the distribution
of subhalos in the Galaxy. We will not present the results of the disant subhalo
calculation of final-state radiation here, since it has been thoroughly explored by
other authors in similar contexts. We direct the interested reader to references
\cite{Kuhlen:2008aw,kuhlen,Anderson:2010df}.

Finally, if a particular subhalo is close enough to subtend an angle
larger than the detector's pixel size, it can no longer be treated as a point source:
eq.\ (\ref{JMH}) must be used, including the angular dependence of the
projected density profile of the given subhalo,
$\rho_{SH}(R,\theta,\phi)$. We will return to this case in Section \ref{sec:closeSub}.

The particle physics contribution to (\ref{phiMH}) and (\ref{phiSH}) comes from the
photon spectrum, defined as:
\begin{equation}
\frac{\d N}{\d \Eg} = \frac{1}{\sv_{\rm total}} \frac{\d\sv}{ \d \Eg}
\end{equation}

\begin{figure}
\hspace{-0.4cm}
\includegraphics[width=0.5\textwidth]{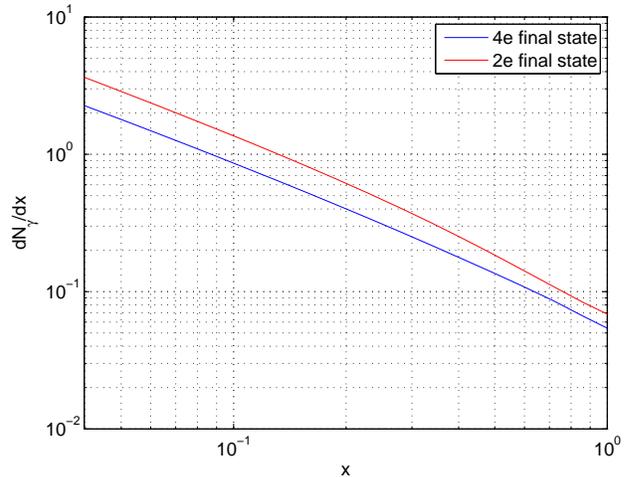}
\caption{Spectrum of prompt gamma rays (brehmsstrahlung) from leptons produced by DM
annihilation, as a function of $x = 2 E_\gamma/\sqrt{s}\cong E_\gamma/M_{DM}$.
 The red line (upper) represents the result of the $2e$ final state, 
and the blue line (lower) corresponds to $4e$ final states.
\label{spectrum1}}
\end{figure}

In the case of a two-lepton final state~\cite{Birkedal:2005ep}:
\begin{equation}
\frac{\d N}{\d x} =\frac{\alpha}{\pi}\frac{1+(1-x)^2}{x} \text{log} \left(\frac{s(1-x)}{m_e^2}\right)
\end{equation}
where $x=2 E_\gamma/\sqrt{s}$ and $s$ is the standard Mandelstam variable. We are
interested in the case of TeV dark matter $\chi$ annihilating to a four-lepton final state,
with a $ \mathcal {O} (1) $ GeV leptophilic gauge boson $B$ as the messenger. 
The annihilation is
dominated by $\chi\chi \to B B$, where the $B$'s are on shell.
The cross section can be obtained by first computing in the rest frame of the $B$
using the decay
$B \rightarrow e^+ + e^-$ and then boosting to the lab frame, in which
the slowly moving DM particles are approximately at rest.  
This can easily be done numerically. We present the resulting spectrum in
fig.~\ref{spectrum1}. Since we will not make use of the final-state bremsstrahlung for 
other annihilation channels (4$\mu$ or 4$\pi$) we will not discuss their spectra. 

\subsection{Inverse Compton Scattering}
\label{sec:ICS}

Charged particles travelling through the interstellar medium scatter off ambient
photons of the interstellar radiation field (ISRF), which is composed of microwave
($\sim 10^{-3}$ eV) radiation from the cosmic microwave background (CMB), infrared
($\sim 10^{-2}$ eV) radiation from dust, and optical ($\sim$ eV) photons from
starlight. Along with the galactic magnetic fields, this is the main source of energy
loss for electrons diffusing within the Galaxy. We will show that ISRF photons that
have scattered with TeV-scale electrons have spectra that peak at several hundred GeV,
which should fall squarely within the measurement window of diffuse gamma rays by the
Fermi Large Area Telescope (LAT).

Once integrated over scattering angles, the well-known Klein-Nishina formula for the
Compton scattering process e$^{\pm}\gamma \rightarrow$ e$^{\pm}\gamma'$ can be
integrated along the line of sight to give the total flux of scattered photons per
solid angle arriving on a detector \cite{Blumenthal:1970gc, Meade}:
 \begin{equation}
 \frac{d\Phi_{\gamma'}}{d\Egp d\Omega} = \frac{1}{2}\hbar^2 c^3 
\alpha_{EM}^2\int_{\rm l.o.s.}\!\!\!\!\!ds \int \int \frac{dn_e}{dE_e}\frac{du_\gamma}{d\Eg}\frac{d\Eg}{\Eg^2}\frac{dE_e}{E_e^2}f_{IC}
\label{loseq}
\end{equation}
$\int_{\rm l.o.s.}ds$ represents the line-of-sight integral from the observer's position 
to infinity (practically speaking, to the edge of the diffusion zone). We have used the 
definitions:
\begin{equation}
 f_{IC} = 2q\log q + (1+2q)(1-q) +\frac{1}{2}\frac{(\epsilon q)^2}{1+\epsilon q}(1-q)
\end{equation}
and
\begin{equation}
 \epsilon = \frac{\Egp}{E_e}, \quad
\Gamma = \frac{4\Eg E_e}{m_e^2}, \quad
q = \frac{\epsilon}{\Gamma(1-\epsilon)}.
\end{equation}

We numerically integrated eq.\ (\ref{loseq}) along the line of sight, as well as over
the incoming particle energies. All the quantities in the integrand are known: we used
the two-dimensional ($r$,$z$) distribution of electrons and positrons $dn_e/dE_e$ from
DM annihilations produced with GALPROP, as discussed in Section \ref{sec:cr}. For the
ISRF, we used a realistic two-dimensional photon energy density distribution
$du_\gamma/dE$ from \cite{Porter:2005qx}, which is publicly available on the GALPROP
website. Both distributions assumed cylindrical symmetry around the Galactic axis. For
each galactic latitude-longitude pair, the line of sight integration was performed in a
three-dimensional sky from the Sun's position to the edge of the diffusion zone which
was taken to extend to a radius $r_{max} = 20$ kpc and to a height $|z|_{max} = 5$ kpc
above and below the galactic plane. A trapezoidal integration step size of 0.1 kpc was
found to be numerically converged. The values of $dn_e/dE_e$ and $du_\gamma/dE$ at each
step were found in the heliocentric coordinate system by using a bilinear interpolation
scheme. On top of the DM annihilation products, we used the densities of primary and
secondary electrons as well as secondary positrons to compute the ICS contribution of
the background lepton field. This had the effect of further constraining the gamma ray
background. 

We performed the integration once per grid point on an equally-spaced $20^\circ\times 
20^\circ$
latitude-longitude grid of the quarter-sky in the ranges $\theta = [0,\pi/2]$, $\phi =
[0,\pi]$. This was sufficient to reconstruct the entire sky, given the symmetry of the
data input.

\subsection{Fermi all-sky diffuse gamma ray measurements}

The Fermi Large Area Telescope (LAT) is a high-sensitivity gamma ray 
instrument capable of detecting photons in the $\sim 30$ MeV to $>$ 300 GeV range. It has 
an effective detector area of $\sim 8000$ cm$^2$, a 2.4 sr field of view and can resolve
the angle of an incident photon to 0.15$^\circ$ at energies above 10 GeV. Data from the
first year of observation are publicly available from the Fermi collaboration.

We used the all-sky diffuse photon file from the Fermi weekly LAT event data webpage
\cite{fermisky}. This covered observations from mission elapsed time (MET) 
239557417 to  MET 272868753 (seconds),
corresponding to 55 weeks of observation between August 8 2008 and August 25 2009. We
processed the photon data with the Fermi LAT science tool software, available from the
Fermi Science Support Center (FSSC) website. We first removed all events with a zenith
angle greater than 105$^\circ$ to eliminate Earth albedo. The data were further trimmed to
keep only the photons measured during ``good'' time intervals. We then created an
exposure cube from the spacecraft data for the corresponding period, to account for
effective instrument exposure. The data were separated into 0.25$^\circ\times 0.25^\circ$
latitude and longitude bins spanning the entire sky, and into 16 logarithmically
separated bins from 100 MeV to 200 GeV. Uncertainties were assigned according to Ref.\
\cite{Porter:2009sg}. We compared our results to the August-December 2008 $10^0 \leq
|b| \leq 20^\circ$ spectrum presented by the Fermi collaboration \cite{Porter:2009sg}. The
half-year data agreed exactly, while adding the extra 8 months to the full 55-week
dataset changed the picture only very slightly. We rebinned the data into a 40$\times$40 grid, 
in correspondance with the ICS computation. 

Before proceeding to the results of our numerical analyses, we should note that many
factors contribute to the theoretical uncertainty. While we were able to reproduce
the results of Simet et al.\ \cite{Simet:2009ne} quite closely, there are substantial
discrepancies between the results of GALPROP and other methods of solving the transport
equation. This lack of agreement is further discussed in \cite{Cline:2010ag}. There is
an additional uncertainty in the injection spectrum of primary electrons, which serve,
along with secondary electrons and positrons from spallation, as the astrophysical
background to our results.

\section{Empirical fits}
\label{sec:results}

As expected, we found that allowing subhalos to contribute to the overall flux of DM
annihilation products reduced the flux of expected gamma rays from the galactic center,
while increasing fluxes at higher galactic latitudes. The most stringent constraints
were from the low-longitude regions just above and just below the galactic plane, where
astrophysical sources of gamma rays are less prominent, but the DM distribution is
still quite dense. Specifically, we used the lower right-hand region ($-9^\circ < b <
-4.5^\circ$, $0^\circ < \ell < 9^\circ$ in Galactic coordinates) which was found to be
the most constraining, in agreement with Ref.\ \cite{Papucci:2009gd}.

After including the ICS from background electrons and positrons, we found that the
boost factor of a main halo 1 TeV DM annihilation process cannot violate the bound $BF
\leq 25$ if the signal is to remain below the top Fermi LAT error bars. If we extend
the constraint to $\Phi_\gamma < $ Fermi $+ 2\sigma$, this condition is only slightly
relaxed to $BF \leq 30$. In the case of a 2.2 TeV DM candidate, these bounds become $BF
\leq 42$ and $BF \leq 52$ at $1 \sigma$ and $2 \sigma$, respectively. While this agrees
qualitatively with other works \cite{Meade,Papucci:2009gd}, we attribute our more
stringent upper bounds mainly to our higher $\rho_\odot$, as discussed in section 
\ref{sec:simulations}, to our inclusion of the ICS contribution from  background 
electrons and positrons,but mainly to the different method used to solve the diffusion 
equation (\ref{diffeq}).

\begin{figure}
\hspace{-0.4cm}
\includegraphics[width=.5\textwidth]{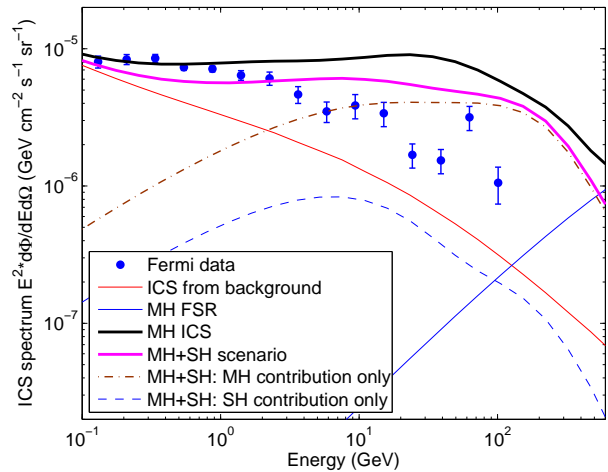}
\caption{Galactic-center ICS gamma ray flux from the region $-9^\circ <  b < -4.5^\circ$, 
$0^\circ < 
\ell < 9^\circ$ for the MH scenario ($M_{DM} = 1$ TeV), top black solid line, are reduced
in the MH+SH scenario ($M_{DM} = 2.2$ TeV), middle magenta solid line, but not enough
to overcome constraints from Fermi LAT observations, which are violated by as much as
4$\sigma$. The parameters for the Einasto profile are $\alpha = 0.17$, $r_s = 25$ kpc.
The background gamma rays (red solid line) include only ICS from background electrons
and positrons, but clearly constrain the model even more. Further contributions are
expected from bremsstrahlung, extragalactic gamma rays and $\pi^0$ decays. The latter
may dominate the spectrum at these energies and are responsible for the hump
shape around 1 GeV \cite{Porter:2008ve}.}
\label{gammaFermi1}
\end{figure}

Using the best fit scenario of Ref.\ \cite{Cline:2010ag}, the reduction of flux was
however not enough to overcome the constraints from the Fermi observations. This is
illustrated in figure \ref{gammaFermi1}, which shows that the MH+SH scenario still
violates constraints from the data by as much as $4 \sigma$. On its own, the predicted
flux exceeded the data at energies above 100 GeV by at least $2\sigma$, while we expect
that additional constraints from $\pi^0\to 2\gamma$ decays should also be large in this 
energy range
\cite{Porter:2008ve} and  push predictions from this model even farther outside
of the observationally allowed region. Allowing the background to freely vary (top
section of Table \ref{resultTable}) made no appreciable difference with respect to
gamma rays, and was not enough to satisfy the observational constraints.

\begin{figure}
\hspace{-0.4cm}
\includegraphics[width=.5\textwidth]{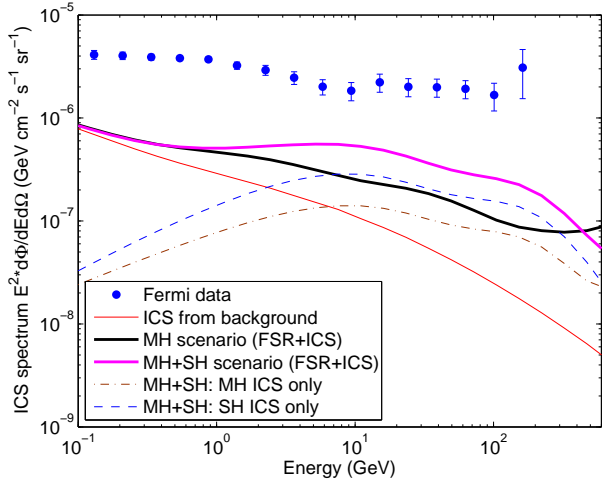}
\caption{Mid-latitude ICS gamma rays from the region 42$^\circ < |b| <
47^\circ$, $9^\circ <|\ell|< 18^\circ$. In this case the MH scenario ($M_{DM} = 1$
TeV), black solid line, predicts fewer ICS gamma rays than
the MH+SH scenario ($M_{DM} = 2.2$ TeV, magenta solid line). 
At these latitudes constraints are much weaker, and neither model is ruled out by the 
observations.}
\label{gammaFermi2}
\end{figure}

Figure \ref{gammaFermi2} illustrates how the ICS gamma ray flux is increased at higher
galactic latitudes when subhalos are included. It should however be emphasized that the
predicted fluxes in this region of the sky are still well below the level of Fermi 
observations. 

\subsection{Less cuspy dark matter profiles}

\begin{figure}
\hspace{-0.4cm}
\includegraphics[width=.5\textwidth]{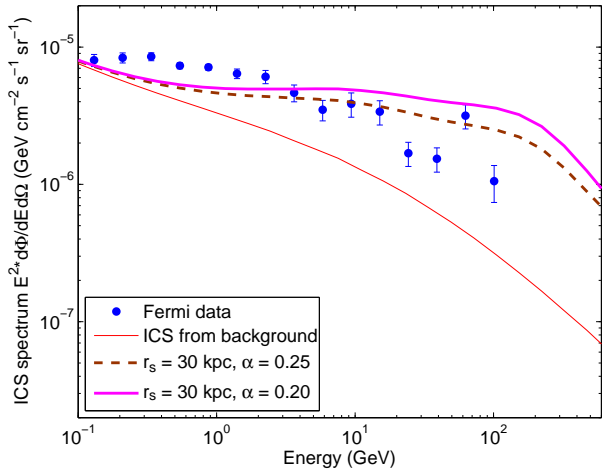}
\caption{ $-9^\circ <  b < -4.5^\circ$, 
$0^\circ < 
\ell < 9^\circ$ region. Similar to previous figures, showing how reducing the cuspiness of 
the Einasto profile (eq.\ (\ref{EinastoEq})) reduces predicted total
gamma ray signal (magenta line). Here $\alpha = 0.20, 0.25$ respectively and
 $r_s = 30$ kpc.}
\label{alpha020}
\end{figure}

\begin{figure}
\hspace{-0.4cm}
\includegraphics[width=.5\textwidth]{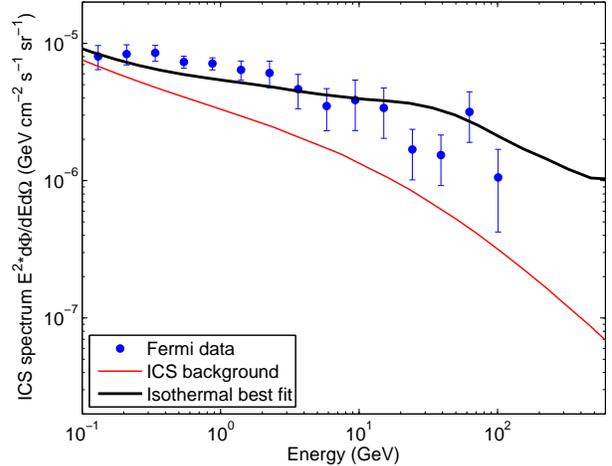}
\caption{$-9^\circ <  b < -4.5^\circ$, 
$0^\circ < 
\ell < 9^\circ$ region. Similar to previous figures, using the cored isothermal profile
with $r_s = 3.2$ kpc and $\rho_s = 3.0$ GeV/cm$^3$.}
\label{isothermal}
\end{figure}

In section \ref{sec:simulations} we mentioned the motivations for
considering less cuspy DM profiles.   Many previous works studying
the ICS constraints have compared the effects of cored versus cuspy
DM profiles, noting that the constraints are weaker for cored
profiles.   To better quantify exactly how much cuspiness can be
tolerated, it is interesting to vary the parameters of the Einasto
profile that control this \cite{Cholis:2009gv,Cirelli:2010nh}. In
particular, larger values of $\alpha$ and $r_s$  correspond to less
concentrated halos.   We ran simulations of the lepton distribution
and gamma ray fluxes with slightly different parameters for equation
(\ref{EinastoEq}) while keeping the local density constant at
$\rho_\odot = 0.37$ GeV cm$^{-3}$. This is illustrated in fig.\
\ref{alpha020}.  Flatter profiles with $\alpha = 0.20$ or $0.25$,
$r_s = 30$ kpc reduce the gamma ray fluxes somewhat, but not enough
to bring the predicted flux to within the observations in the
offending energy bins between 10 and 100 GeV.   The same is true for
the isothermal profile, whose corresponding results are shown in
fig.\ \ref{isothermal}.  For both cases, the problem arises because
the predicted background gamma flux is not far below the observed
flux in the most constraining bins.  This leaves very little room for
the additional contribution from the DM decay products ICS signal.

Increasing the intermediate gauge boson mass to 1 GeV, and thus
allowing a decay to muons and pions according to the branching ratios
described in Section \ref{sec:channels} does not alleviate the
problem. Indeed, the 1$\sigma$ (2$\sigma$) bounds become BF $< 23$
($<28$) for an Einasto profile, and BF $< 63$ ($< 72$) in the
isothermal case. These fall well short of the required BF = 118 to
explain the Fermi and PAMELA excesses, as long as the DM mass is
increased to $M_{DM} = 1.2$ TeV. These results are summarized in the
bottom of Table \ref{subhaloTable}. The reason ICS constraints are stronger when muons are included is due to the nature of the data. Indeed, the peak of the ICS spectrum lines up with the most constraining data point when $M_{DM} = 1.2$ TeV. This provides a stronger than expected constraint, relative to the 4e final state at $M_{DM} = 1$ TeV. 

\begin{table}
 \begin{tabular}{|c|c|c|c|c|c|}
\hline
 Subhalo& $r_s$ (kpc) & $\rho_s$  & $\log BF$ & $d_{\rm min}$ (pc) &
 $V_{\rm max}$ (km/s)\\ \hline
1 & 0.01 & 69 & 4.74 & 33.9 & 2.9\\
2 & 0.1 & 3.46 & 4.34 & 95.5 & 6.7\\
3 & 3.2 & 0.04 & 3.76 & 178 & 22 \\
4 & 0.9 & 1.27 & 2.35 & 165 & 36 \\  
5 & 1.1 & 2.0  & 1.70 & 170 & 55 \\
\hline
\multicolumn{6}{|c|}{Main halo, 4e channel} \\
\hline
Einasto & 25 & 0.048 & $\phantom{\Big|}<{1.40\atop 1.48}$ & $-$ & $201-277$\\
Isothermal & 3.2 & 2.32 & $\phantom{\Big|}<{1.81\atop 1.88}$ & $-$ & $201-277$ \\
\hline 
\multicolumn{6}{|c|}{Main halo, 4e + 4$\mu$ + 4$\pi$ channel} \\
\hline
Einasto & 25 & 0.048 & $\phantom{\Big|}<{1.36\atop 1.45}$ & $-$ & $201-277$\\
Isothermal & 3.2 & 2.32 & $\phantom{\Big|}<{1.80\atop 1.86}$ & $-$ & $201-277$ \\
\hline
\end{tabular}
\caption{Upper rows: parameters of each subhalo we examined. $r_s$ and $\rho_s$ 
(in units GeV cm$^{-3}$)
characterize the halo's Einasto profile (with $\alpha=0.17$), 
$\log BF$ is the logarithm of the necessary boost factor in order to obtain the 
Fermi lepton data entirely from the given subhalo
and $d_{\rm min}$ is the minimum distance (in pc) from our 
position to such a subhalo along the sun-GC axis, with the given boost factor, 
that would not exceed the gamma ray 
observations.  $V_{\rm max}$ is the maximum circular velocity, which appears
in the radial velocity dispersion, fig.\ \ref{vl2vel}.
Lower rows: similar data for the main halo using Einasto or isothermal profiles,
but $\log BF$ denotes the  
1 and $2\sigma$ upper limits
to satisfy gamma ray constraints. 
}
\label{subhaloTable}
\end{table}

\begin{figure} \hspace{-0.4cm} \includegraphics[width=.5\textwidth]{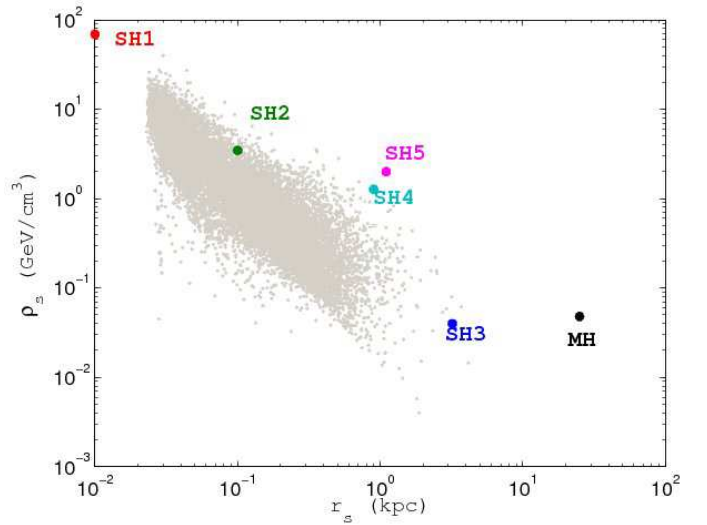}
\caption{Grey regions: scatter plot of $\rho_s$ versus $r_s$ for subhalos
in the \vl\ simulation. Dots represent the main halo (MH) and subhalos given in table
\ref{subhaloTable}.}
\label{shdist} \end{figure}

\begin{figure} \hspace{-0.4cm} \includegraphics[width=.5\textwidth]{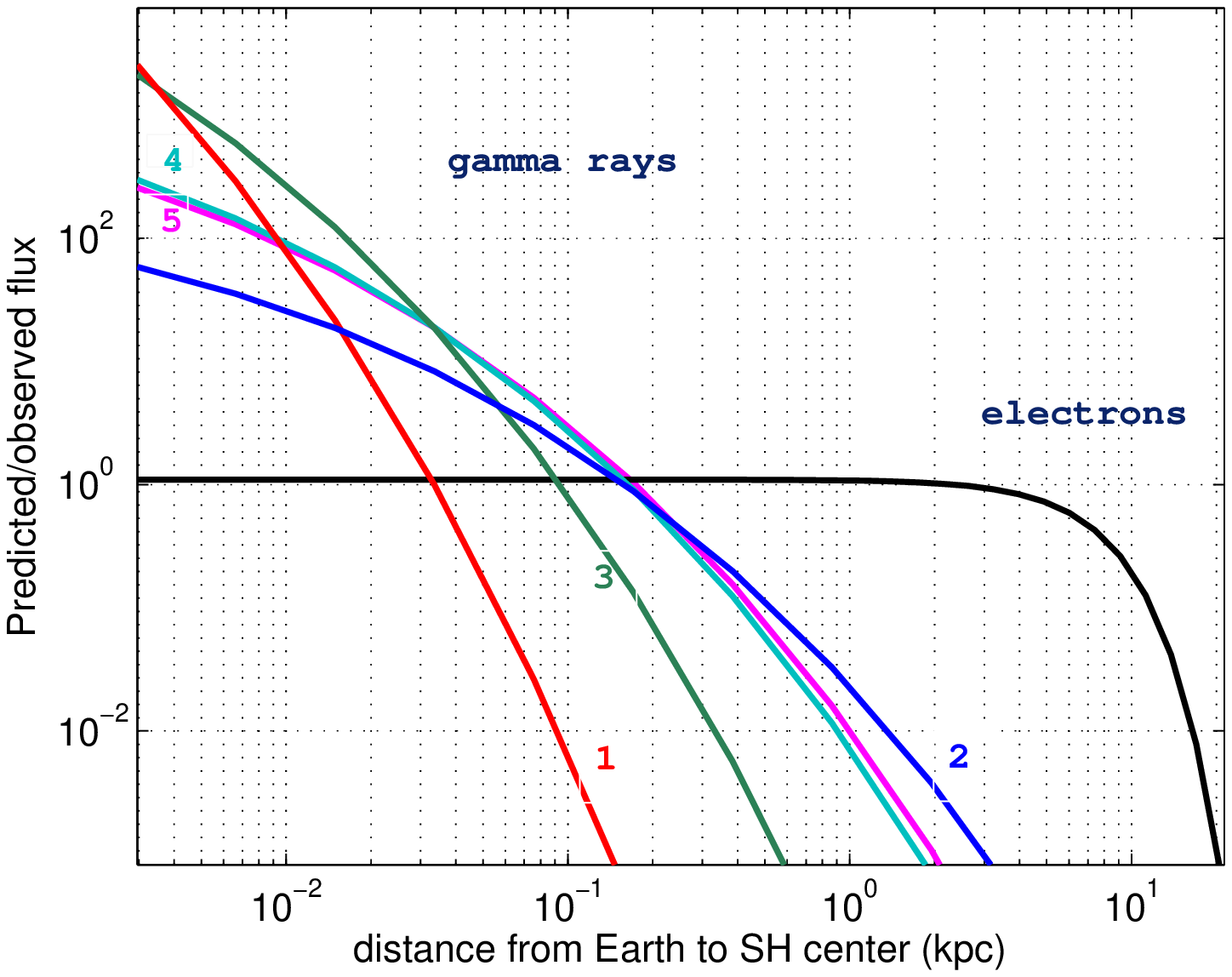}
\caption{Fluxes of gamma rays  and $e^+ + e^-$  from the five subhalos
presented in Table \ref{subhaloTable}. The gamma ray fluxes (curve
labeled by the number of the corresponding subhalo) are at $E_\gamma = 137$ GeV,
whereas the leptons are at an energy of 559 GeV (the peak of the
observed Fermi spectrum). In both cases, the amplitude
is the predicted flux divided by the observed flux from the Fermi satellite, such that
a value of $10^0$ means that the predicted flux is equal to the observed value. Boost
factors in each case (as given in table \ref{subhaloTable}) were fixed to allow the 
Fermi lepton signal to be explained
entirely by the subhalo. The allowed position of each subhalo with respect to earth
is therefore the region to the right of each gamma ray curve, up to $\sim 2$ kpc
where the lepton flux starts to fall.}
\label{subhaloConstraints} \end{figure}

\subsection{Close subhalo}
\label{sec:closeSub}

The above analyses implicitly assume that no single subhalo dominates the lepton
signal. But if a subhalo happens to be very close (within a kpc) to the solar 
system,  the picture changes
significantly, since the electrons and positrons from the close subhalo can
 dominate the observed flux, and its gamma ray emissions can come from a sizable
solid angle in the sky. We treat this case separately from the previous subhalo
scenario, since a larger DM mass is no longer required to produce the observed lepton
signal; rather, the small amount of ICS energy loss during propagation from a local
subhalo means that a 1 TeV-scale DM particle appropriately conforms to the Fermi $e^+ +
e^-$ measurements. We concentrate on the 4e final state channel, although previous
results allow this to be generalized. The solution depends linearly
on the spectrum $dN_e/dE$, so that the boost factor required to
explain the observed lepton excess should scale in the same way that
it does in the main halo scenario: $BF_{(e,\mu,\pi)}/BF_{4e}
\simeq 118/110$, as read from Table \ref{resultTable}.

Since GALPROP is not easily adapted in its 2D mode to include the effects of a highly 
localized 
additional source term, we adopt a semi-analytic approach to solve the diffusion 
equation (\ref{diffeq}) for 
leptons produced in the nearby subhalo. Given that the leptons and gamma rays in this 
scenario
would be from a local origin, the spatial dependence of the interstellar radiation and
magnetic fields becomes much less important. We used the method described in ref.\
\cite{Baltz:1998xv}, with the same diffusion parameters as presented in section
\ref{sec:cr} (of the present work), but with an energy-loss coefficient parametrized 
by
\begin{equation}
 b(x,E)=-\frac{\d E_e}{\d t} = \frac{E_e^2}{\tau_E}
\end{equation} 
with $\tau_E = 10^{16}$ s GeV characterizing the local energy loss rate.

We sampled subhalos from the \vl\ simulation to identify examples that could allow
for simultaneously fitting the PAMELA/Fermi lepton fluxes and the Fermi gamma
ray fluxes.  Four such examples are labeled as SH1-SH4 in table \ref{subhaloTable},
and a fifth (SH5) is one that we have ``engineered'' by choosing parameters that
are close to those of SH4, but with a higher density and hence higher circular
velocity, dynamically related to each other by eq.\ (13) of
\cite{kuhlen}, 
\be
V_{\rm max}^2 = f_V 4\pi G\rho_s r_s^2
\label{vmax}
\ee
with $f_V = 0.897$.  Due to the higher
density, SH5 requires a lower boost factor to produce the observed lepton signal,
and so it represents a kind of best-case scenario.  The distribution
of \vl\ subhalos in the space of $(r_s,\rho_s)$ is shown as a scatter plot in fig.\ 
\ref{shdist},
and the five subhalos of interest are highlighted on this plot.  They are atypical
in the sense of needing a higher-than-average central density. A further caveat is that such a large $r_s$ is unlikely at small distances from the GC due to tidal disruption. Indeed, subhalos within the visible galaxy in the \vl  simulation were of the order $r_s = 0.05 \sim 0.85$ kpc, falling below the $0.9 \sim 1.1$ kpc compatible with the most plausible particle physics scenario discussed in Section \ref{sec:particle}.

Each subhalo was situated along an optimal axis, namely that connecting the earth to
the GC.  Such an accidental alignment makes it easier to ``hide'' the gamma rays 
originating from the subhalo since they are coming  primarily from the same direction
as the GC, where the background emissions are strongest.  This is also the reason
that the most stringent ICS constraints on the main halo arise from the regions
$4.5^\circ < |b| < 9^\circ$ of galactic longitude instead of the most central region. 
However in this case we find that the biggest contribution to the emission is from
final-state bremsstrahlung rather than ICS.  The latter is found to produce gamma ray
fluxes that are 3 orders of magnitude smaller than observed. This is consistent with
the fact that the main source of ICS is IR radiation and starlight, which is
concentrated far from the vicinity of  the solar system.   

Results were then compared to the Fermi lepton and gamma ray data in order to establish
constraints. The strictest gamma constraints were at the largest energy data point from
the Fermi LAT analysis of $E = 162$ GeV, because of the shape of the FSR spectrum, which
 rises steadily until $\sim$ 1 TeV. We used a slightly different region of the sky than in
our previous ICS analysis, $4.5^\circ < |b| < 9^\circ$, $9^\circ < |\ell| < 18^\circ$,
because there were not enough good data points in this energy bin at lower longitudes
to constrain the data. We compared the lepton
prediction to the Fermi measurements at 559 GeV, where the observed $e^++e^-$
spectrum is at a maximum deviation from a power law. In both cases we included the
additional constraints from astrophysical backgrounds computed by GALPROP and by our
ICS routine. 

Results are shown in fig.\ \ref{subhaloConstraints}. If the single
subhalo is allowed to saturate the observed lepton signal, fig.\
\ref{subhaloConstraints} gives clear bounds (summarized in Table
\ref{subhaloTable}) on the proximity of each subhalo, providing a
minimum distance from the solar neighborhhod to such a subhalo.  So
long as the boost factor for the main halo remains sufficiently
small, this scenario can therefore overcome the ICS constraints that
restricted the standard MH-only model.

\subsection{Astrophysical prediction and extragalactic constraints}

\begin{figure}
\hspace{-0.4cm}
\includegraphics[width=.5\textwidth]{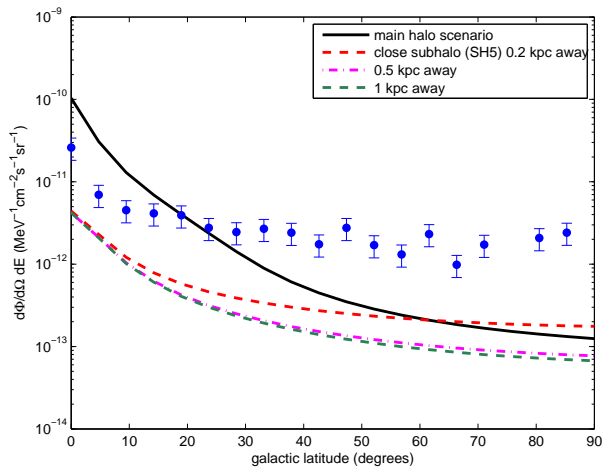}
\caption{Dependence of predicted gamma ray fluxes on galactic 
latitude $b$, in the region $-9^\circ < \ell < 9^\circ$ at 
E = 23 GeV, the most constraining energy bin for the main halo 
scenario. Black: main halo scenario (Einasto profile, BF = 110) 
Dashed: subhalo 5, as specified in Table \ref{subhaloTable}. 
Background ICS is included in both predictions, but signal is
dominated by final state radiation. Dots are the Fermi 
data for that region and energy.}
\label{latpredictions}
\end{figure}

In figure \ref{latpredictions} we provide an example of the gamma ray
flux predicted by the close subhalo scenario, as compared to the main
halo scenario.  The gamma ray flux comes predominantly from final state
radiation rather than inverse Compton scattering of the annihilation
products.  We chose the energy bin E = 23 GeV, which is the most
constraining for the main halo case. Although both scenarios converge
at high latitudes, low latitude measurements have already ruled out
the main halo scenario, and provide a way to constrain the model.
With more exposure and precise removal of point sources, the Fermi
LAT may provide a diffuse background low enough to rule out these
predictions. As a further test, census experiments such as the
upcoming Gaia satellite may provide a precise enough map of the local
gravitational potential to confirm or rule out the presence of such a
DM overdensity  \cite{Brown:2008sh}. Direct measurement of such an
overdensity would however be difficult: a subhalo such as SH5,
located at a distance that would not saturate gamma ray bounds, would
contribute less than $0.1 \%$ of the local DM density.

From previous works, we infer that extragalactic bounds on this scenario are not as strong as the ones we have computed above. Bounds from dwarf spheroidal galaxies could plausibly be important since the velocity dispersions are of the same order as what is required for our subhalo enhancement, i.e. $\sim  10-50$ km s$^-1$ \cite{Walker:2007ju}. However, the most stringent Fermi LAT bounds \cite{Abdo:2010ex} from such galaxies put the upper limit on DM annihilation into a 2$\mu$ final state at around $BF = 3000$ if only final-state radiation is considered, and around 300 if ICS bounds are included as well. \cite{Abdo:2010dk} computed the cosmological dark matter annihilation bounds for the same 2$\mu$ final state scenario, and find that $BF$ larger than 300 is excluded at the 90\%  confidence level. This is using the results of the Millennium II structure formation simulation, and is indeed model-dependent. Extrapolation to the 4$\mu$ scenario is independent of astrophysics. We can therefore take the results of \cite{Meade,Papucci:2009gd} who have construced bounds on both channels. They show that FSR bounds are consistently an order of magnitude weaker in the 4$\mu$ case, given the softer photon spectrum in this scenario. We can therefore take these extragalactic results to be far less constraining than the stringent bounds from the center of our own galaxy.

\begin{figure}
\hspace{-0.4cm}
\includegraphics[width=.5\textwidth]{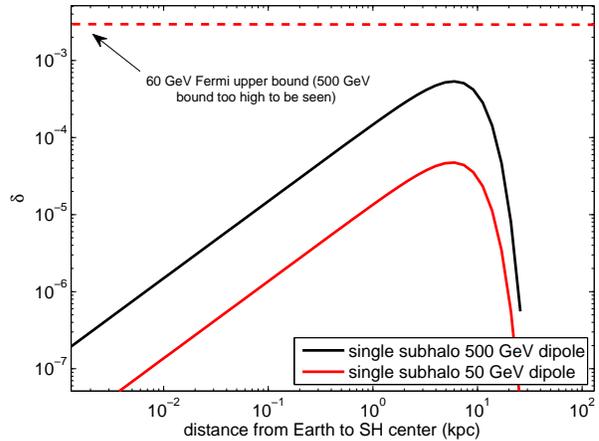}
\caption{Dipole anisotropy $\delta$ of the cosmic ray electron and positron flux predicted by SH5 if it saturates the Fermi excess. Background cosmic ray electrons and positrons are included, and taken to be isotropic. $\delta$ increases monotonically with energy from the red line (60 GeV) to the black line (500 GeV). }
\label{dipolefig}
\end{figure}

Finally, we verify that this model does not saturate bounds on dipole anisotropy of the cosmic ray $e^+ + e^-$ spectrum. The dipole anisotropy can be defined as 
\begin{equation}
 \delta = 3\sqrt{\frac{C_1}{4\pi}},
\end{equation} 
where $C_1$ is the standard dipole power of the measured electron and positron flux in the sky. The Fermi LAT collaboration \cite{Ackermann:2010ip} have presented upper bounds on this quantity. These range from $\delta \lsim 3 \times 10^{-3}$ at $E_e \simeq 60$ GeV up to  $\delta \lsim 9 \times 10^{-2}$ at $E_e \simeq 500$ GeV. Given a diffusive model, this can be computed \cite{Ackermann:2010ip}:
\begin{equation}
 \delta = \frac{3D(E)}{c}\frac{|\vec \nabla n_e|}{n_e},
\end{equation} 
where $D(E)$ is the diffusion coefficient (\ref{diffusioncoefficient}) and $n_e$ is the density of cosmic ray electrons and positrons, including astrophysical backgrounds. Taking the background to be isotropic, we computed the dipole anisotropy in the case of a single close subhalo producing enough electrons to explain the Fermi excess. In every case $\delta$ falls well below bounds. Results for SH5 are presented in Figure \ref{dipolefig}. The anisotropy rises monotonically with energy, from 60 GeV (red line) to 500 GeV (black line).

\section{Particle physics realizations}
\label{sec:particle}

In the previous sections we have identified scenarios where subhalos
could provide the observed excess PAMELA and Fermi leptons, from a
purely phenomenological perspective.  In particular,  certain values
for the annihilation cross section boost factors are needed for the
subhalos, and upper bounds for that of the main halo (depending upon
assumptions about its density profile) were derived.  It is interesting
to ask whether simple particle physics models with boost factors from
Sommerfeld enhancement can be consistent with these requirements.  

The simplest possibility for model building is dark matter that
annihilates into light scalar or vector bosons, which subsequently
decay into leptons.  This class of models automatically gives a 
boost factor to the annihilation cross section, through multiple
exchange of the boson, resulting in Sommerfeld enhancement.  However
it is not obvious that one can find models with the desired boost
factors for the subhalos and main halo.  One constraint that limits
our freedom is to not exceed the measured density of dark matter. It
will turn out that our mechanism works most naturally if the DM
responsible for signals in the galaxy is a subdominant component
comprising some fraction $1/f$ of the total DM population 
\cite{Cirelli:2010nh}, with $f>1$. 

We focus on the case of a GeV-scale U(1) vector boson that kinetically 
mixes with the photon. Such models have the advantage of naturally 
explaining the coupling to light leptons, without producing excess
antiprotons that would contradict PAMELA observations. Let us denote
the vector's mass by $\mu$ and the coupling by $g$, with $\alpha_g = g^2/4\pi$. 
If $M$ is the DM mass, then the Sommerfeld boost factor is controlled
by two dimensionless parameters: $\epsilon_\phi = \mu/(\alpha_g M)$ and
$\epsilon_v =  v/(\alpha_g c)$, where $v$ is the DM velocity in the
center of mass frame.  A reasonably accurate approximation to the exact
Sommerfeld enhancement is given by the expression 
\cite{Cassel:2009wt,Slatyer:2009vg} \be S = {\pi\over \epsilon_v}{\sinh
X \over \cosh X - \cos\sqrt{{2\pi\over\bar\epsilon_\phi} - X^2}}
\label{Seq} \ee where $\bar\epsilon_\phi= (\pi/12)\epsilon_\phi$ and $X
= \epsilon_v/\bar\epsilon_\phi$.  (The cosine becomes $\cosh$ if the
square root becomes imaginary.)

To take into account leptophilic DM that is only a subdominant
component of the total DM,  suppose that $\alpha_{g,\rm th}$ is the
value of $\alpha_g$ that would give the correct thermal abundance,
which scales like the inverse annihilation cross section $\sigma^{-1}
\propto \alpha_g^{-2}$; then we can parametrize $\alpha_g = \sqrt{f}\,
\alpha_{g,\rm th}$.  The rate of annihilations goes like
$\rho_l^2\sigma \propto 1/f$ if $\rho_l$ stands for the
leptophilic component of the DM.  We accordingly
define an effective boost factor  
\be \bar S = {S\over f} \ee
 where $S$ is the intrinsic Sommerfeld enhancement factor.  Thus any
constraint on $S$ in a theory with $f=1$ becomes a constraint on 
$\bar S$ in the more general situation.

\begin{figure}[t] \hspace{-0.4cm}
\includegraphics[width=.5\textwidth]{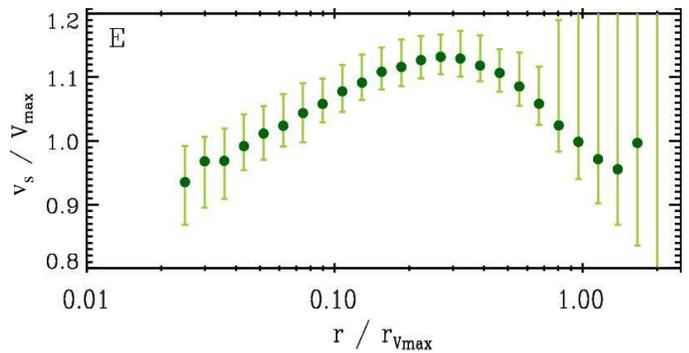} 
\caption{Radial velocity dispersion of subhalos in the \vl\ simulation,
taken from ref.\ \cite{kuhlen}} \label{vl2vel} \end{figure}

\subsection{Averaging of boost factor}
\label{bavg}

Of course, the DM velocity has no definite value; instead we need to 
average over the possible values within the subhalos and the main
halo, weighted by the appropriate distribution function.  We take
it to be Maxwell-Boltzmann with a cutoff at some escape velocity,
\be
	f(v) = N e^{-3v^2/2 v_s^2}\, \theta(v_{\rm esc} -v)
\ee
This isotropic form is only an approximation since the true
distribution has some small anisotropy between the radial and angular
components; we will for simplicity ignore this complication.
The velocity dispersion $v_s = \langle v^2\rangle^{1/2}$ depends upon 
the radial distance $r$
from the center of the halo or subhalo.  The dependence has been
measured for the subhalos in the \vl\ simulation; see figure
\ref{vl2vel}.  The shape is universal, but is scaled along the
respective axes by parameters $V_{\rm max}$ and $r_{V_{max}}$ that depend
upon the subhalo.  The latter is related to the scale radius
by $r_{V_{max}} = 2.212\, r_s$; the former is given by (\ref{vmax})
and also listed in table 
\ref{subhaloTable} for the subhalos of interest.  For numerical
purposes we fit the sides of the curve passing through the 
points of fig.\ \ref{vl2vel} by lines (omitting the rightmost point),
and the middle by an inverted parabola.\footnote{The velocity
dispersion curve is fit by 
\vskip-0.5cm
\be
y =\left\{\begin{array}{ll} 1.309 + 0.232 x,& x < -0.841,\\
 0.976 - 0.3437x,& x > -0.383 \\
0.9618 - 0.5475x - 0.4413x^2,& \hbox{in between}\end{array}
\right.
\ee
where $x=\log_{10} r/r_{V_{\rm max}}$ and $y = v_s/V_{\rm max}$.}
   We use the same form of 
$v_s$ for the main halo, with $r_s = 25$ kpc and $V_{\rm max} = 
201$ km/s.  Other authors have advocated higher values of the 
velocity dispersion, $v_s = 309$ km/s at $r=r_\odot$ 
\cite{Savage:2009mk}, which would correspond to $V_{\rm max} = 
277$ km/s in the present parametrization.  We will also consider the
higher value to take account of this uncertainty.

The escape velocity can be computed explicitly for the subhalos from
the standard result $\frac12 v_{\rm esc}^2 = G\int_r^\infty 
(M(r)/r^2)\,dr$, where $M(r)  = 4\pi\int_0^r r^2 \rho\, dr$ is the mass within
radius $r$.  The result for an Einasto profile is
\bqa
	v_{\rm esc}^2 &=& G\,\rho_s\, e^{2/\alpha}\,{8\pi\over\alpha}\left(
	{\alpha\over 2}\right)^{3/\alpha}\left[
\left(\sfrac{2}{\alpha}\right)^{1/\alpha}
\Gamma\left(\sfrac{2}{\alpha}, \sfrac{2}{\alpha}(\sfrac{r}{r_s})^\alpha\right) 
\right.\nonumber\\
&+& \left.\frac{r_s}{r}\left(\Gamma\left(\sfrac{3}{\alpha}\right) -
	\Gamma\left(\sfrac{3}{\alpha}, \sfrac{2}{\alpha}
(\sfrac{r}{r_s})^\alpha\right)\right) \right],
\eqa
where $\Gamma(s,x)$ is the upper incomplete gamma function. For the main halo, 
this procedure would not be correct because of the
significant contribution of baryons, not included here.  We adopt the
result for $v_{\rm esc}$ of ref.\ \cite{Cirelli:2010nh} 
for the main halo (see appendix C of that reference).

With these ingredients, we can compute an average Sommerfeld
enhancement factor $\langle S\rangle$ for each subhalo:
\be
	\langle S\rangle ={ \int_{r_1}^{r_2} dr\, r^2\, \rho^2 \int d^3 v_1\,  d^3 v_2\,
	f(v_1)\, f(v_2) S(\frac12|\vec v_1-\vec v_2|) \over 
	\int_{r_1}^{r_2} dr\, r^2\, \rho^2}
\ee
The factor of $\sfrac12$ in the argument of $S$ occurs because the $v$ appearing in eq.\
(\ref{Seq}) through $\epsilon_v$ is half of the relative velocity.
$\rho^2$ is the appropriate weighting factor because the rate of
annihilations is proportional to $\langle\sigma v\rangle \rho^2$.
For the subhalos, the range of integration for $r$ is from 0 to
$\infty$, but for the main halo we take lower and upper limits
$r_{1,2}$ that correspond to the angular region of the sky that is used
to set the gamma ray constraints: $r_1 = 0.67$ kpc and $r_2 = 1.34$
kpc.  The reason is that the bound $\bar S< 30$ for the main halo
comes from the gamma ray constraint rather than from lepton
production.  We are thus interested in the boost factor relevant to
the region $4.5^\circ < |b| < 9^\circ$ of galactic latitude.  The
distances of closest approach to the galactic center, hence largest
rate of $\gamma$ ray production associated with
these lines of sight, are given by $r = r_\odot \sin b$.

\begin{figure}[t] \hspace{-0.4cm}
\includegraphics[width=.5\textwidth]{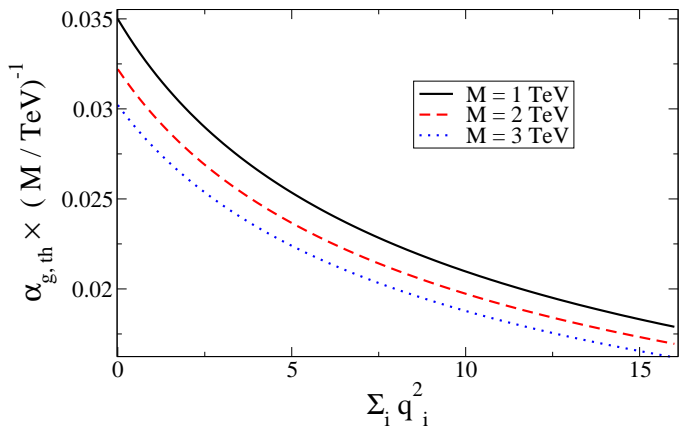} 
\caption{Value of gauge coupling leading to correct thermal relic DM
density, $\alpha_{g,\rm th} /M$, versus squared charge of dark Higgs 
bosons in U(1) model,
for several values of DM mass $M$.} \label{alphath} \end{figure}

\subsection{Relic Density Constraint}
The enhancement factor (\ref{Seq}) depends rather strongly on the 
gauge coupling $\alpha_g$; therefore it is interesting to know what
constraint the relic density places upon $\alpha_g$. The effect of a Sommerfeld-enhanced DM model on the relic densitie has been discussed by \cite{Zavala:2009mi}.
Notice that DM
transforming under a U(1) gauge symmetry as we have assumed must be
Dirac and therefore could have a relic density through its asymmetry,
similar to baryons.   However, unless the DM was never in thermal
equilibrium, then $\alpha_g$ should not be less than the usual value
$\alpha_{g,\rm th}$ leading to the correct relic density, since otherwise 
the thermal component will be too large.

There are two kinds of final states for annihilation of DM in this
class of models: into a pair of gauge bosons $B_\mu$, by virtual DM exchange
in the $t$ and $u$ channels, or into dark Higgs bosons $h$, by exchange
of a gauge boson in the $s$ channel.  Assuming the DM ($\chi$) is much heavier
than the final states, the respective squared amplitudes, averaged
over initial and summed over final spins, are
\be
	\frac14\sum|{\cal M}|^2 = \left\{\begin{array}{ll}
	4 g^4(1 + 2 v^2), & \chi\chi\to B B \\
	\frac12 g^4 q^2 (1 - v^2\cos^2\theta),& \chi\chi\to h\bar h
	\end{array} \right.\\ \phantom{A}
\ee
where $q$ is the U(1) charge of $h$ relative to $\chi$ (replace $q^2\to
\sum_i q_i^2$ for multiple Higgs bosons), $\theta$ is the scattering
angle, and we have included the leading dependence on the initial velocity
$v$ in the center of mass frame.  The factor $\cos^2\theta$ averages
to $2/3$ in the integral over $\theta$.  
In computing the associated cross
section, it must be remembered that the $2B$ final state consists of identical
particles, while the Higgs channel does not.  The total amplitude can
therefore be written in the form $\frac14\sum|{\cal M}|^2 = 
4g^4(a + bv^2)$, with
\be
	a = 1 + \sfrac14\sum_i q_i^2,\quad 
	b = 2(1 - \sfrac{1}{12} \sum_i q_i^2)
\ee
if we use the phase space for identical particles.

To find the cross section relevant during freeze-out in the early
universe, we thermally average the $v$-dependent $\sigma v_{\rm rel}$
following ref.\ \cite{Cline:2010kv}.  We include approximately the effect of
Sommerfeld enhancement as described there, to obtain
\bqa
\langle\sigma v_{\rm rel}\rangle &\cong& {\pi\alpha_g^2\over 2 M^2}
\left( a\left(1 + \alpha_g\sqrt{\pi \sfrac{M}{T}}\right)\right. \nonumber \\ 
    &+& \left. {T\over M}(b-\sfrac43 a)
	\left(\sfrac32 + \alpha_g\sqrt{\pi\sfrac{M}{T}}\right) \right)
\eqa
The terms that are subleading in $\alpha_g$, but enhanced by
$\sqrt{M/T}$, are due to the Sommerfeld correction.  We approximate
the freezeout temperature as $T\cong M/20$, the usual result of
solving the Boltzmann equation for DM in the TeV mass range, and equate
$\langle\sigma v_{\rm rel}\rangle$ to the value 
$\langle\sigma v\rangle_0 = 3\times
10^{-26}$ cm$^3$/s usually assumed to give the correct relic density.
This gives an implicit equation for $\alpha_{g,\rm th}$ of the form
$\alpha_g^2 = c_1 M^2 \langle\sigma v\rangle_0/( 1 + c_2 \alpha_g)$, which however quickly
converges by numerically iterating.  Fig.\ \ref{alphath} displays the resulting
dependence of $\alpha_{g,\rm th}/M$ on $\sum_i q_i^2$ for several values of $M$.

The bound that the density of the leptophilic DM component not exceed
the total DM density is $\alpha_g > \alpha_{g,\rm th}$.  We
parametrize the coupling by $\alpha_g = \sqrt{f}\,\alpha_{g,\rm th}$ with $f>1$
in what follows.

\begin{figure}[t] \hspace{-0.4cm}
\includegraphics[width=.5\textwidth]{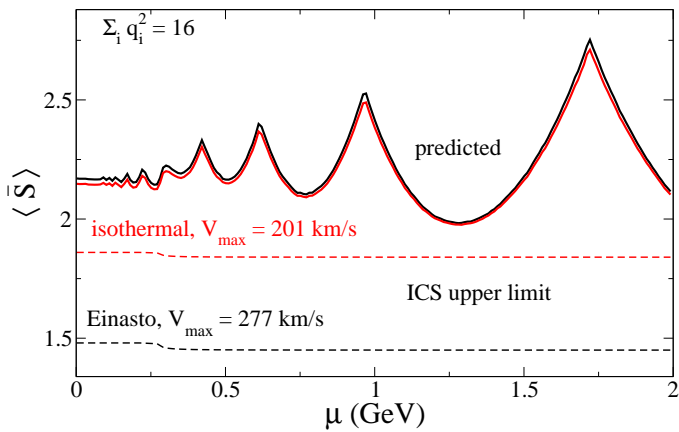} 
\caption{Solid lines: predicted main halo boost factor for 
thermal value of $\alpha_g$, with dark Higgs boson charges $\sum_i
q_i^2 = 16$ and maximum circular velocity $V_{\rm max} = 277$ km/s.
Upper curve is for Einasto profile, lower for isothermal.
Dashed line is $2\sigma$ upper limit from gamma rays produced by
inverse Compton scattering. The failure to satisfy this bound even
with large dark Higgs content and large $V_{\rm max}$ drives us
to consider larger than thermal gauge couplings, $f>1$.   
} \label{qs16} \end{figure}

\subsection{Interpolation between $4e$ and mixed final states}

In our numerical computations with GALPROP, we considered two cases
for the final state annihilation channels: either $\chi\chi\to 4e$,
applicable for gauge bosons with mass $\mu < 2 m_\mu$, 
or to a mixture of electrons, muons and charged pions, appropriate for
decays of gauge bosons with mass greater than $2 m_\pi$.   The
relative abundance of $e$, $\mu$ and $\pi$ in the mixed final state
can be computed from the branching fractions of the decays, 
discussed in connection with eq.\ 
(\ref{branchratios}).

For intermediate values of the gauge boson mass, $2 m_\mu < \mu \lsim
2 m_\pi$, we can use the branching ratios to interpolate between our 
maximum-allowed MH or best-fit SH boost factors for the $4e$ case and
those of the fiducial $e+\mu+\pi$ case.  The maximum allowed boost
factors of the main halo complying with the ICS constraints are taken
from table II.  To estimate the best fit boost factors for the
subhalos in the fiducial $e+\mu+\pi$ final state, we rescale the 
$4e$ results shown in table II by the ratio of best-fit boost factors
for the main halo, in the MH-only scenario.   These ratios are
$118/110$ for the Einasto profile and $119/113$ for the isothermal,
quite close to unity, and so the best-fit values of the
SH boost factors hardly depend upon this scaling.   More significant
is the change in the best-fit mass, from $M=1.0$ to 1.2 TeV, which enters
into the computation of the Sommerfeld enhancement and the value of
the gauge coupling ($\alpha_g\sim M$).  We use the branching ratios to
interpolate $M$ as well. For the MH upper bounds in the small- and large-$\mu$ regions, we use the values from Table II, and interpolate similarly for intermediate $\mu$.

\begin{figure*}[t]
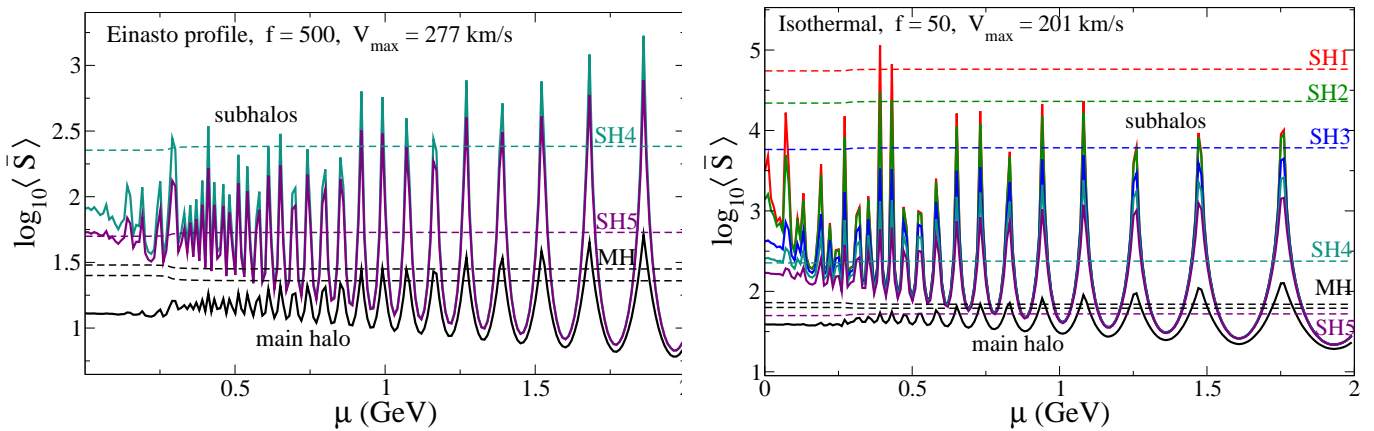
 \hspace{-0.4cm}
\centerline{
\includegraphics[width=.51\textwidth]{f500-ein.eps} \hfil
\includegraphics[width=.49\textwidth]{f50-iso-new.eps}
}
\caption{Predicted effective boost factors $\langle \bar S\rangle$
as a function of gauge boson mass
(solid curves) and target values (or upper limit in case of main
halo, dashed curves) to explain PAMELA/Fermi lepton observations
and Fermi gamma ray constraints.  Pair of dashed curves for main halo
(MH) correspond respectively to 1 and 2$\sigma$ upper limits.  
Left panel is for $f=100$, $V_{\rm
max} = 277$ km/s,  Einasto main halo profile; 
right is for $f=25$, $V_{\rm, max} = 201$ km/s, isothermal main halo
profile.  Subhalos are those of table \ref{subhaloTable}.  Points which
satisfy all constraints are those where subhalo curves intersect
their corresponding dashed line while the main halo curve lies below
its dashed lines.
} \label{working} \end{figure*}

\subsection{Theoretical fits}

For a given value of the gauge coupling $\alpha_g$, we can determine
the predicted boost factors as well as the desired values for each
subhalo, as a function of the gauge boson mass $\mu$, and similarly for
the main halo, except here we have an upper bound on $\langle \bar S\rangle$
rather than a best-fit value.  This bound in fact presents the biggest
challenge to finding a working particle physics model.  For $\alpha_g$
close to the thermal relic density value $\alpha_{g,\rm th}$, the
predicted boost factor of the main halo far exceeds the bound $\langle 
\bar S\rangle \lsim 30$, even if we try to decrease
$\langle  \bar S\rangle$ by reducing $\alpha_g$ via a
large hidden Higgs content  or by increasing the 
dispersion of the main halo.  Fig.\ \ref{qs16}
illustrates the discrepancy for $\sum_i q_i^2 = 16$ and $V_{\rm max} =
277$ km/s.  Lower values of $V_{\rm max}$ or 
$\sum_i q_i^2$ only make this tension worse.

As we mentioned above, even though it is not theoretically possible
to make the gauge coupling weak enough to solve this problem,
ironically one can rescue the scenario by {\it increasing} $\alpha_g$
beyond the thermal value, since this suppresses the relic density of
the DM component we are interested in, and thus reduces the
scattering rate.  Allowing $\alpha_g = \sqrt{f}\alpha_{g,\rm th}$
decreases both the density of the leptophilic component and  the
effective boost factor by $1/f$.  With $f\sim 50-500$, depending upon
the shape of the main halo DM density profile,  we can satisfy the
constraint on the MH and still have a large enough boost in certain
hypothetical nearby subhalos for them to supply the observed lepton
excess.  The minimum value of $f$ that is needed is larger for a cuspy
main halo.

We give two working examples in figure \ref{working}, one with
$f=500$ and $V_{\rm max} = 277$ km/s (the larger value advocated in
ref.\ \cite{Savage:2009mk}) and assuming an Einasto profile for the
main halo, and the other having $f=50$ and $V_{\rm max} = 201$ km/s
(the more standard assumption for the velocity dispersion), with an
isothermal halo.  In these figures the averaged boost factor $\langle
\bar S\rangle$ of the relevant subhalos are plotted as solid lines,
while the required values of $\langle \bar S\rangle$ are the dashed
curves.  Wherever these intersect represents a possible value of the
gauge boson mass to consistently explain the observed lepton excess. 
At the same time, the main halo boost factor (lowest solid curve in
the small-$\mu$ region) must lie below the black dashed lines to
satifsy gamma ray constraints.  The rationale for taking the larger
value of $V_{\rm max}$ for the Einasto profile is that larger
velocities help to suppress the boost factors and thus make it easier
to satisfy the ICS constraint, so that we are not forced to choose an
even larger value of $f$.  The isothermal profile is less
constrained.

In the first panel of fig.\ \ref{working} with the Einasto profile,
only subhalos SH4 and SH5 have large enough boost factors to ever
reach the required values.  There are many points of intersection, 
but mainly those for SH5 and in the mass range $\mu < 750$ MeV are
consistent with the gamma ray bounds on the main halo.  For the
isothermal halo, these constraints are less stringent, and it is
possible to find points of intersection using $f=50$ for all five of
the sample subhalos, although they are much more rare for SH1$-$SH3
than for SH4 and SH5.  In this example, the intersection points that
respect the ICS bound are restricted to $\mu\lsim 1$ GeV.  For larger
values of  $f$, all the boost factors will be further suppressed, and
$\mu > 1$ GeV will become allowed for SH4 and SH5.  

\begin{figure}[t] \hspace{-0.4cm}
\includegraphics[width=.5\textwidth]{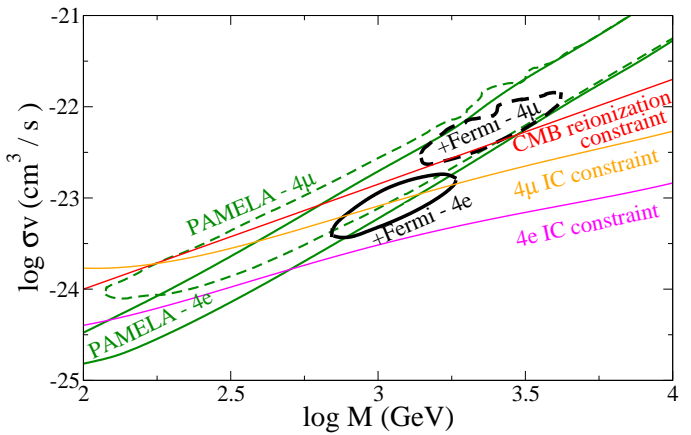} 
\caption{Allowed regions for PAMELA and Fermi excess leptons, and
upper bounds from inverse Compton gamma rays, from ref.\ 
\cite{Papucci:2009gd}, for Einasto profile with $\alpha=0.17$ and
$r_s= 20$ kpc.  CMB constraint is from ref.\ 
\cite{Cirelli:2009bb}.
} \label{spbounds} \end{figure}

One advantage of
requiring large $f$ is that the corresponding dilution of the DM
density by $1/f$ insures that the model satisfies stringent CMB constraints
from annihilations in the early universe changing the optical depth
\cite{Huetsi:2009ex,Cirelli:2009bb,Slatyer:2009yq}, as pointed out in 
\cite{Cirelli:2010nh}.  The CMB constraint is shown in fig.\
\ref{spbounds}, along with the PAMELA/Fermi allowed regions from 
ref.\ \cite{Papucci:2009gd} for $4e$ and $4\mu$ final states. 
The $4e$ case is allowed by the CMB constraint, but $4\mu$ is 
ruled out.  Because our model has at most a fraction of $0.45$ of
muons in the final state, it is probably already safe, 
but the additional weakening of the bound by the factor $1/f$ ensures
that this will be the case.  Similarly, our scenario overcomes the
no-go result of ref.\ \cite{Feng:2010zp}, which pointed out that 
Sommerfeld enhanced annihilation in the early universe leads to
constraints on the MH boost factor which are lower than those
needed to explain the lepton anomalies.  Our MH boost factor can
satisfy these constraints since the MH is no longer considered to
be the source of the excess leptons.

The Sommerfeld enhancement is nearly saturated for the  low velocities
of the subhalos at these large values of $\alpha_g\sim 0.1-0.35$, so
their $\langle\bar S\rangle$ curves are nearly overlapping except at
the smallest gauge boson masses.  The main halo boost factor is not
saturated on the other hand, and lies below the FSR bound for most
values of $\mu$.  We have chosen the gauge couplings, parametrized by
$f$, to nearly saturate the FSR bound.  By taking larger $\alpha_g$
(larger $f$), the bounds could be satisfied by a larger margin.  But 
this would also reduce the $\langle\bar S\rangle$ values of the subhalos
by a similar amount, making it more difficult to get a large enough
lepton signal from SH1$-$SH3.  SH4 and SH5 would remain robust possible
explanations.

\section{Discussion and Conclusions}
\label{sec:conclusion}

We have shown that gamma ray constraints on leptophilic annihilating
dark matter are significantly stronger than in previous studies, when
we take into account the contributions to inverse Compton scattering
from primary and secondary electrons and positrons, before
including  excess leptons from the DM annihilation. We attribute part of this difference to the method of solving the diffusion equation (1) --- fully numerical rather than semi-analytic --- meaning that the $(r,z)$ space-dependence of the diffusion coefficient is taken into account. The difference
between the predicted and observed  spectra of  gamma rays is greatly
reduced, leaving little room for new contributions.  Because of this,
even cored halos, which were allowed by other analyses, become
excluded.  However, we find that these constraints can be  weakened
and possibly overcome if annihilations in a nearby subhalo are the
dominant source of anomalous leptons, rather than annihilations in
the main galactic DM halo. In this way, the PAMELA/Fermi cosmic ray
excesses can be explained, without violating bounds from the recent
Fermi LAT diffuse gamma ray survey. 

It must be admitted that the subhalo loophole we present is rather
special.  First, only atypically dense subhalos, relative to the
sample provided by \vl, give a large enough boost factor (see fig.\
\ref{shdist}).  Second, the subhalo would need to accidentally line
up nearly with the galactic center in order for the ICS gamma rays
associated with these leptons to be sufficiently hidden by the noisy
background of the GC.  Of course, had we
neglected ICS contributions of background electrons, similar to 
previous studies, less fine tuning of the subhalo properties would be
necessary.  Also we do not require the subhalo to be particularly
close; fig.\ \ref{subhaloConstraints} shows that the lepton flux only
starts to fall at distances of $\sim 3$ kpc.  
Our finding could be regarded as a proof of concept.  It
is possible that the effects of unresolved substructure within the
subhalo \cite{Bovy}, which can increase the boost factor, would also
make the scenario work more easily.  On the positive side, there is
the opportunity of testing whether there is such a nearby subhalo,
since we predict the spectrum of ICS gamma rays it contributes (see
fig.\ \ref{latpredictions}).  A better understanding of backgrounds,
for example from point sources, could make it possible to rule out
the proposal.

On the particle physics side, we have shown in detail that the
subhalo scenario can be made consistent with one of the simplest
models of leptophilic dark matter, where the DM is in a hidden sector
that communicates with the standard model only through kinetic mixing
with hypercharge of a new gauge boson in the GeV mass range.  The
relative couplings to leptons and charged pions are completely 
specified and the model has only two free parameters, the gauge
coupling $\alpha_g$ and gauge boson mass $\mu$ (the DM mass $M$ is
fixed by fitting to the spectrum of anomalous $e^++e^-$).  The gauge
coupling is constrained by the relic density of the DM.  The
Sommerfeld enhancement factor is completely fixed by $(\alpha_g, \mu,
M)$ and the kinematical halo properties.  We find (similarly to ref.\
\cite{Cirelli:2010nh}) that the predicted boost factor for the main
halo is always {\it too large} to satisfy ICS constraints unless the
leptophilic component of the DM is small, comprising a fraction of
order $1/f = 0.02-0.002$ of the total DM.  The small fraction can be 
achieved by assuming $\alpha_g$ is larger than the value required 
for the usual thermal abundance by the factor $f\sim 50-500$.  This
raises the interesting possibility that the DM that may be responsible for the
cosmic ray anomalies is distinct from the dominant DM species that
might be discovered by direct detection.  

\section*{Acknowledgments}   

We would like to thank Ilias Cholis for kindly giving us access to his modifications 
to the GALPROP code, and Troy Porter for valuable help with the Fermi
data analysis. We thank Andrew Frey, Thomas Konstandin, Guy Moore,
Natalia Toro and Mike Trott for helpful discussions.
Our research is supported by NSERC (Canada) and FQRNT (Qu\'ebec).

\bibliography{galprop}

\begin{thebibliography}{10}

\bibitem{ArkaniHamed:2008qn}
N.~Arkani-Hamed, D.~P. Finkbeiner, T.~R. Slatyer, and N.~Weiner,
\newblock Phys. Rev. {\bf D79}, 015014 (2009), 0810.0713.

\bibitem{Pospelov:2008jd}
M.~Pospelov and A.~Ritz,
\newblock Phys. Lett. {\bf B671}, 391 (2009), 0810.1502.

\bibitem{Cholis:2008qq}
I.~Cholis, D.~P. Finkbeiner, L.~Goodenough, and N.~Weiner,
\newblock (2008), 0810.5344.

\bibitem{Cholis:2008wq}
I.~Cholis, G.~Dobler, D.~P. Finkbeiner, L.~Goodenough, and N.~Weiner,
\newblock (2008), 0811.3641.

\bibitem{Meade}
P.~Meade, M.~Papucci, A.~Strumia, and T.~Volansky,
\newblock (2009), 0905.0480.

\bibitem{Chen:2009ab}
F.~Chen, J.~M. Cline, and A.~R. Frey,
\newblock (2009), 0907.4746.

\bibitem{Watson}
G.~Kane, R.~Lu, and S.~Watson,
\newblock (2009), 0906.4765.

\bibitem{Kuhlen:2008aw}
M.~Kuhlen, J.~Diemand, and P.~Madau,
\newblock (2008), 0805.4416.

\bibitem{Cirelli:2008pk}
M.~Cirelli, M.~Kadastik, M.~Raidal, and A.~Strumia,
\newblock Nucl. Phys. {\bf B813}, 1 (2009), 0809.2409.

\bibitem{Feng:2010zp}
J.~L. Feng, M.~Kaplinghat, and H.-B. Yu,
\newblock (2010), 1005.4678.

\bibitem{Cline:2010ag}
J.~M. Cline, A.~C. Vincent, and W.~Xue,
\newblock Phys. Rev. {\bf D81}, 083512 (2010), 1001.5399.

\bibitem{Cirelli:2009dv}
M.~Cirelli, P.~Panci, and P.~D. Serpico,
\newblock (2009), 0912.0663.

\bibitem{Strumia:2010zz}
A.~Strumia,
\newblock Prog. Theor. Phys. Suppl. {\bf 180}, 128 (2010).

\bibitem{Cirelli:2010nh}
M.~Cirelli and J.~M. Cline,
\newblock (2010), 1005.1779.

\bibitem{Bovy}
J.~Bovy,
\newblock (2009), 0903.0413.

\bibitem{Kuhlen:2009jv}
M.~Kuhlen,
\newblock (2009), 0906.1822.

\bibitem{Kistler:2009xf}
M.~D. Kistler and J.~M. Siegal-Gaskins,
\newblock (2009), 0909.0519.

\bibitem{Ando:2009fp}
S.~Ando,
\newblock Phys. Rev. {\bf D80}, 023520 (2009), 0903.4685.

\bibitem{kuhlen}
M.~Kuhlen, P.~Madau, and J.~Silk,
\newblock (2009), 0907.0005.

\bibitem{Hutsi:2010ai}
G.~Hutsi, A.~Hektor, and M.~Raidal,
\newblock JCAP {\bf 1007}, 008 (2010), 1004.2036.

\bibitem{Kuhlen:2009is}
M.~Kuhlen and D.~Malyshev,
\newblock Phys. Rev. {\bf D79}, 123517 (2009), 0904.3378.

\bibitem{Brun:2009aj}
P.~Brun, T.~Delahaye, J.~Diemand, S.~Profumo, and P.~Salati,
\newblock Phys. Rev. {\bf D80}, 035023 (2009), 0904.0812.

\bibitem{Strong:1998pw}
A.~W. Strong and I.~V. Moskalenko,
\newblock Astrophys. J. {\bf 509}, 212 (1998), astro-ph/9807150.

\bibitem{Simet:2009ne}
M.~Simet and D.~Hooper,
\newblock JCAP {\bf 0908}, 003 (2009), 0904.2398.

\bibitem{GALPROP:web}
galprop.stanford.edu.

\bibitem{Porter:2005qx}
T.~A. Porter and A.~W. Strong,
\newblock (2005), astro-ph/0507119.

\bibitem{Delahaye:2008ua}
T.~Delahaye {\em et~al.},
\newblock Astron. Astrophys. {\bf 501}, 821 (2009), 0809.5268.

\bibitem{vl2}
J.~Diemand {\em et~al.},
\newblock Nature {\bf 454}, 735 (2008), 0805.1244.

\bibitem{aquarius}
V.~Springel {\em et~al.},
\newblock Mon. Not. Roy. Astron. Soc. {\bf 391}, 1685 (2008), 0809.0898.

\bibitem{Catena:2009mf}
R.~Catena and P.~Ullio,
\newblock (2009), 0907.0018.

\bibitem{Salucci:2010pz}
P.~Salucci,
\newblock (2010), 1008.4344.

\bibitem{Salucci:2000ps}
P.~Salucci and A.~Burkert,
\newblock Astrophys. J. {\bf 537}, L9 (2000), astro-ph/0004397.

\bibitem{Papucci:2009gd}
M.~Papucci and A.~Strumia,
\newblock (2009), 0912.0742.

\bibitem{Cholis:2008vb}
I.~Cholis, L.~Goodenough, and N.~Weiner,
\newblock Phys. Rev. {\bf D79}, 123505 (2009), 0802.2922.

\bibitem{Anderson:2010df}
B.~Anderson, M.~Kuhlen, R.~Johnson, P.~Madau, and J.~Diemand,
\newblock Astrophys. J. {\bf 718}, 899 (2010), 1006.1628.

\bibitem{Birkedal:2005ep}
A.~Birkedal, K.~T. Matchev, M.~Perelstein, and A.~Spray,
\newblock (2005), hep-ph/0507194.

\bibitem{Blumenthal:1970gc}
G.~R. Blumenthal and R.~J. Gould,
\newblock Rev. Mod. Phys. {\bf 42}, 237 (1970).

\bibitem{fermisky}
http://fermi.gsfc.nasa.gov/cgi bin/ssc/LAT/WeeklyFiles.cgi.

\bibitem{Porter:2009sg}
T.~A. Porter and f.~t. F.~L. Collaboration,
\newblock (2009), 0907.0294.

\bibitem{Porter:2008ve}
T.~A. Porter, I.~V. Moskalenko, A.~W. Strong, E.~Orlando, and L.~Bouchet,
\newblock Astrophys. J. {\bf 682}, 400 (2008), 0804.1774.

\bibitem{Cholis:2009gv}
I.~Cholis {\em et~al.},
\newblock (2009), 0907.3953.

\bibitem{Baltz:1998xv}
E.~A. Baltz and J.~Edsjo,
\newblock Phys. Rev. {\bf D59}, 023511 (1998), astro-ph/9808243.

\bibitem{Brown:2008sh}
A.~G.~A. Brown,
\newblock AIP Conf. Proc. {\bf 1082}, 209 (2008), 0810.5437.

\bibitem{Walker:2007ju}
M.~G. Walker {\em et~al.},
\newblock (2007), 0708.0010.

\bibitem{Abdo:2010ex}
A.~A. Abdo {\em et~al.},
\newblock Astrophys. J. {\bf 712}, 147 (2010), 1001.4531.

\bibitem{Abdo:2010dk}
Fermi-LAT, A.~A. Abdo {\em et~al.},
\newblock JCAP {\bf 1004}, 014 (2010), 1002.4415.

\bibitem{Ackermann:2010ip}
Fermi LAT, M.~Ackermann {\em et~al.},
\newblock (2010), 1008.5119.

\bibitem{Cassel:2009wt}
S.~Cassel,
\newblock J. Phys. {\bf G37}, 105009 (2010), 0903.5307.

\bibitem{Slatyer:2009vg}
T.~R. Slatyer,
\newblock JCAP {\bf 1002}, 028 (2010), 0910.5713.

\bibitem{Savage:2009mk}
C.~Savage, K.~Freese, P.~Gondolo, and D.~Spolyar,
\newblock JCAP {\bf 0909}, 036 (2009), 0901.2713.

\bibitem{Zavala:2009mi}
J.~Zavala, M.~Vogelsberger, and S.~D.~M. White,
\newblock Phys. Rev. {\bf D81}, 083502 (2010), 0910.5221.

\bibitem{Cline:2010kv}
J.~M. Cline, A.~R. Frey, and F.~Chen,
\newblock (2010), 1008.1784.

\bibitem{Cirelli:2009bb}
M.~Cirelli, F.~Iocco, and P.~Panci,
\newblock JCAP {\bf 0910}, 009 (2009), 0907.0719.

\bibitem{Slatyer:2009yq}
T.~R. Slatyer, N.~Padmanabhan, and D.~P. Finkbeiner,
\newblock Phys. Rev. {\bf D80}, 043526 (2009), 0906.1197.

\bibitem{Huetsi:2009ex}
G.~Huetsi, A.~Hektor, and M.~Raidal,
\newblock (2009), 0906.4550.

\end{thebibliography}
 \bibliographystyle{h-physrev}
\end{document}